\documentclass[structabstract]{aa}  

\usepackage{graphicx}
\usepackage{lscape}
\usepackage{txfonts}
\begin{document}
   \title{Self-consistent nonspherical isothermal halos embedding zero-thickness disks}

   \subtitle{}

   \author{N.C. Amorisco
          \inst{1}
          \and
         G. Bertin\inst{2}
          }

   \institute{Institute of Astronomy, Madingley Road, Cambridge CB3 0HA, UK\\
             \email{amorisco@ast.cam.ac.uk}
         \and
Dipartimento di Fisica, Universit\`a degli Studi di Milano,
via Celoria 16, I-20133 Milano, Italy\\
              \email{giuseppe.bertin@unimi.it}
                          }

   \date{Received March 10, 2010; accepted May 8, 2010}

 
  \abstract
   {That the rotation curves of spiral galaxies are
generally flat in the outer regions is commonly considered an
indication that galaxy disks are embedded in
quasi-isothermal halos. In practice, disk-halo decompositions of
galaxy rotation curves are performed in a parametric way by
modeling the halo force contribution by means of expressions
describing approximately the properties of the regular isothermal
\textit{sphere} or other \textit{spherical} density distributions suggested by
cosmological simulations.}
   {We construct self-consistent models of \textit{nonspherical} isothermal halos
embedding a zero-thickness disk, by assuming that the halo
distribution function is a Maxwellian. The method
developed here can be used to study other
physically-based choices for the halo distribution function and 
the case of a disk accompanied by a bulge.}
   {The construction is performed by means of an iterative procedure that generalizes a
method introduced in the past. In a preliminary investigation we note the existence of a fine tuning between
the scalelengths $R_{\Omega}$ and $h$, respectively characterizing the rise of
the rotation curve and the luminosity profile of the disk, which surprisingly applies  
to both high surface brightness and low surface
brightness galaxies. This empirical correlation identifies a 
much stronger conspiracy than the one required by the smoothness and flatness 
of the rotation curve (\textit{disk-halo conspiracy}).}
   {The self-consistent models are found to be characterized by smooth and flat rotation curves for very different disk-to-halo mass ratios, hence suggesting that conspiracy is not as dramatic as often imagined. For a typical rotation curve, with asymptotically flat rotation
curve at $V_{\infty}$ (the precise value of which can also be treated as a free parameter), 
and a typical density profile of the disk,
self-consistent models are characterized by two
dimensionless parameters, which correspond to the dimensional
scales (the disk mass-to-light ratio $M/L$ and the halo central
density) of standard disk-halo decompositions. We show that if
the rotation curve is decomposed by means of our self-consistent
models, the disk-halo degeneracy is removed and typical
rotation curves are fitted by models that are below the
maximum-disk prescription. Similar results are obtained from a study of NGC 3198. Finally, we quantify the flattening of the spheroidal halo, which is significant, especially on the
scale of the visible disk ("abridged").}
   {}


   \maketitle

%

\section{Introduction}

\label{Sect:Introduction}

The study of the rotation curve of spiral galaxies (from the case of M31, van de Hulst et al. 1957, to the systematic investigation of nearby galaxies in the 70's) has led to convincing evidence for the presence of dark matter halos (e.g., see Roberts 1976). To this discovery, HI rotation curves, which often sample regions well beyond the bright optical disk, played a decisive role (van Albada et al. 1985; van Albada \& Sancisi 1986).

In this context, it has been suggested that the existence of a halo of dark matter introduces a problem of fine tuning or ``conspiracy" (van Albada \& Sancisi 1986). The coexisting dark and visible components, while expected to have different formation and evolution histories (even if naturally coupled by their mutual gravitational interaction), must always cooperate to produce a gravitational field inversely proportional to the radius, so that the resulting rotation curve remains flat from the inner regions (generally thought to be dominated by the visible matter) out to large radii (where the field is dominated by the diffuse dark halo). This puzzling feature is accompanied by a serious problem of {\it degeneracy}. In fact, from a detailed study of the so-called disk-halo decomposition, it has been shown that, in general, the available photometric profile and the corresponding velocity data-points can be modeled in many different ways, either by heavy or by light disks, through an ill-determined mass-to-light ratio $M/L$ for the disk. In general, the available data are only able to set an upper limit to such $M/L$, identifying the so-called maximum-disk solution, for the model that maximizes the contribution of visible matter. Typically, the maximum-disk solution is able to fit the rotation curve out to the optical radius (Kalnajs 1983; Kent 1986).

It was immediately realized that the flatness of the outer rotation curves suggests that the dark halo embedding the visible disk might have a structure similar to that of the classical isothermal sphere, because outside the central core such sphere is characterized by a density distribution declining approximately as $\rho \sim 1/r^2$. Curiously, the ubiquitous presence of gravitational fields consistent with a flat rotation curve also came from studies of elliptical galaxies, not only in the nearby universe (for NGC 4472, see lower left frame of Fig.~7 of Saglia et al. 1992 and Bertin 1993; for IC 2006, see Franx et al. 1994; for a more recent systematic study, see Gerhard et al. 2001), but also for ellipticals at cosmologically significant distances (e.g., see Barnab\'e et al. 2009, and references therein).

Of course, deviations from the paradigmatic case of galaxies such as NGC 3198 are not uncommon (e.g., see Casertano \& van Gorkom 1991). But it has been pointed out that such deviations are generally associated with well-defined features present in the distribution of visible matter or the presence of non-circular motions (Sancisi 2004). On the other hand, pure disks characterized by the absence of specific photometric features are quite common (see also Kautsch 2009).

The literature on the general topic of the distribution of dark matter in spiral galaxies is enormous and deals with a variety of important aspects of the problem. In recent years these studies have generally focused on the requirements of current cosmological scenarios. In this Introduction it would be impossible to do justice to all the relevant studies in this general research area. However, it appears that one simple aspect of the problem has gone practically unnoticed (but see some discussion given by van Albada \& Sancisi 1986). This is the issue of the mutual interactions between halo and visible matter in determining the self-consistent equilibrium of the basic state. In their review, van Albada \& Sancisi (1986) addressed this point, with the help of N-body simulations, but did not carry out a systematic investigation.

Here we wish to reconsider this problem, starting from an extremely simple physical (and somewhat academic) description of the dark halo. We will thus address the problem of determining the nonspherical structure of a dark halo, assumed to be perfectly isothermal, in the presence of a given disk. We will show that even if very simple, and thus lacking some important features present in real objects (for example, the presence and role of gas are not considered explicitly in this paper, although they could be easily incorporated), this study leads to some interesting and unexpected results.

The search for equilibrium solutions of the Vlasov-Poisson equations, where the mean potential is generated by the distribution of stars populating the system, with the possible contribution of an ``external" density distribution (e.g., associated with a co-existing dark matter halo), is a classical problem of stellar dynamics, with important analogues in the study of collisionless plasmas. In this paper we follow the approach based on the Jeans theorem to deal with the halo matter, while the disk of visible matter is taken to provide an external field. In this approach, the starting point is the identification of an appropriate form for the one-particle distribution function in phase space; in our case this is the isothermal assumption for the dark halo. The adopted distribution function then generates a $\Phi$-dependent density in physical space, thus leading, inserted as a source in the Poisson equation, to a nonlinear problem in $\Phi$. Here $\Phi$ represents the gravitational potential. In the past, this method has been used frequently and has led to important insights into the dynamics of galaxies and other stellar systems. Yet the study of self-consistent equilibria in the presence of external fields that break the natural symmetry associated with the one-component problem is only rarely considered (see Bertin \& Varri 2008). Therefore, besides the natural application to the discussion of the structure of dark halos around galaxies, this paper has some interest of its own, as a study in stellar dynamics.

The structure of the paper is the following. In Sect.~2 we comment on the so-called disk-halo conspiracy. Here we identify an empirical correlation which brings out a new and stronger conspiracy than the one associated with the flatness of rotation curves. In Sect.~3 we formulate the self-consistent problem of constructing an isothermal spheroid in the presence of a zero-thickness disk. The actual procedure to calculate the self-consistent models is reported in Appendix A. The properties of self-consistent isothermal halos are then described in Sect.~ 4. Then in Sect.~5, by modeling first a fiducial rotation curve and later by modeling the specific case of NGC 3198, we show that self-consistency actually removes the degeneracy that characterizes standard parametric disk-halo decompositions (with additional comments given separately in Appendix B). Finally, Sect.~6 summarizes the main results of the paper. This article is all based on the mathematically simple case of zero-thickness disks. A follow-up paper will address the corresponding study of disks with finite thickness.

\section{A comment on the disk-halo ``conspiracy"}
\label{Sect:Conspiracy}
 
The rotation curve $V(R)$ of many spiral galaxies is featureless,
regular, and smooth: after a central region of approximately
linear growth, the rotation velocity reaches a constant value $V_{\infty}$ and
remains flat out to large radii, often well beyond the bright optical
disk. [But, of course, this is meant to be only a zero-th order description, because even flat rotaton curves are not entirely featureless (Sancisi \& van Albada 1987).] Since the underlying gravitational field results from the
sum of the contributions of different mass components of the
system (the stellar disk, the bulge, the gaseous disk, and the
dark matter halo), which generally dominate in different regions,
the smoothness and lack of features of the rotation curve is
unexpected and poses a physical problem, which is
commonly referred to as the \textit{disk-halo conspiracy}.  

When such conspiracy was noted (see van Albada \& Sancisi 1986) most
of the emphasis was placed on the flatness of the rotation
curves. In reality, it has long been noted that standard parametric disk-halo decompositions, based on the
superposition of the fields produced by the visible matter and by
a \textit{spherical halo}, are subject to a high level of degeneracy. 
Equally viable fits to the rotation curves are obtained with very different
models (i.e. very different values of the mass-to-light ratio of
the stellar disk, assumed to be constant within the galaxy), ranging from some characterized by a very light
disk and a dominant halo all the way in, to others, close to the
so-called maximum disk solution, where the support to the rotation
curve inside the optical disk is basically due to the visible
matter alone, while the halo contribution becomes important and
eventually dominates only outside. In a sense, the existence of such degeneracy
actually undermines, or at least softens, the problem of conspiracy apparently raised by the flatness of rotation curves.

Of course, there are several constraints that can be used to reduce the level of degeneracy present. In addition to those posed by stellar
population analyses, we may recall the constraints posed by dynamical stability in the context of spiral structure (Ostriker \& Peebles 1973; Bertin \& Lin 1996; see also Widrow et al. 2008; and references therein) or warp dynamics (e.g., see Bertin \& Mark 1980).

To give a concrete example of the procedure adopted in standard
parametric disk-halo decompositions, we may recall the case of a
purely exponential disk with the adoption of the model of a classical
pseudo-isothermal \textit{sphere} for the halo contribution. In this
case the rotation curve is decomposed into $V_{mod}^2=V_{DM}^2+V_D^2$.
Here 
\begin{equation}
V^2_D(\xi)= {V_{\infty}^2}{\beta\over
8}\xi^2\left[I_0\left({\xi\over2}\right)K_0\left({\xi\over2}\right)-I_1\left({\xi\over2}\right)K_1\left({\xi\over2}\right)\right]\ ,
\label{vd}
\end{equation}\noindent where $\xi=R/h$ is the galactocentric cylindrical radius in units of the exponential scalelength
and $\beta$ is a dimensionless parameter (see Eq.~(\ref{beta})) which measures the weight of the stellar disk.
In addition, $I_0$, $K_0$, $I_1$, and $K_1$ denote the standard modified Bessel functions. The halo contribution is described by
\begin{equation}
V_{DM}^2(\xi)=
V^2_{\infty}\left[1-\sqrt{2\over\alpha}{1\over\xi}\arctan\left(\sqrt{\alpha\over
2}\xi\right)\right]\ , 
\label{vdm}
\end{equation}
\noindent where $\alpha$ is a dimensionless measure of the central dark matter density (see Eq. (\ref{alpha})). In this example, the degeneracy occurs in the $(\alpha, \beta)$ parameter space (see Sect.~\ref{decomp}).
 
In the following subsection we will show the existence of a much stronger reason for invoking a problem of conspiracy, associated with the fine tuning between the length scale that characterizes the inner
growth of the rotation curve and the natural scale length of the visible disk 
(in the absence of significant contributions from a
bulge or from gas, this would be the exponential length). This tuning is a direct indication of a subtle interplay between the density distribution of dark and visible matter. If the two length scales were not sufficiently close to the observed correlation, the degeneracy of the standard parametric disk-halo decompositions would not take place.

\subsection{An important empirical correlation} \label{empiric}
 
This subsection will provide a useful empirical background to the
physically simple dynamical models presented in this paper. Here
we start by identifying one important aspect of the problem of the
uniformity of the observed characteristics of rotation curves.
 
For simplicity, we refer to a sample of galaxies for which the
contribution of the bulge should be negligible. We first consider
disk galaxies belonging to the Ursa Major cluster, for which
good-quality photometric profiles (Tully et al. (1996)) and rotation
curves (Verheijen \& Sancisi (2001)) are available. Without actually aiming at
producing a complete sample in any statistically significant
sense, which would go well beyond the goals of this paper, we then
add a small number of other galaxies, with data taken from
Kent (1987), de Blok \& McGaugh (1997) (with photometries from de Blok et al. (1995)),
 and Swaters et al. (2000) (with photometries from de Blok et al. (1995) and from McGaugh \& Bothun (1994)), so as to increase
the size of the sample. The only criterion used to select the
galaxies from the source papers is the shape of their photometric
profiles. Galaxies with clear deviations from an exponential
profile in the central parts, often related to the presence of a
bulge, are discarded. Therefore, for these approximately exponential stellar
disks, any difference in the shape of the rotation curves should
be interpreted only in terms of the different relative weights of
the disk and of the dark component (under the simplifying assumption that the mass-to-light ratio is constant within each galaxy disk). The final sample (see Table 1)
thus consists of thirty-five objects, almost equally divided
between High Surface Brightness and Low Surface Brightness
galaxies. 
 
For each galaxy we consider the exponential length $h$ of the
stellar disk and the radius $R_{\Omega}$ that we define as the
radius at which the rotation curve reaches two thirds of its estimated
asymptotic flat value $\hat V_{\infty}$:
 
\begin{equation}
V(R_{\Omega})={2\over 3}\hat V_{\infty}\ .
\label{romega}
\end{equation}
 
\noindent The velocity $\hat V_{\infty}$ is the estimate of the asymptotic velocity 
$V_{\infty}$ obtained by averaging the rotation velocity in the outer parts of the rotation curve 
(see also Sect.~\ref{flat}). The radius $R_{\Omega}$ thus measures the slope of the inner
approximately linear growth of the rotation curve.
 
\begin{table}

\begin{center}
\begin{tabular}{|c|c|c|c|c|}\hline

\textbf{Name}&\textbf{Type}&$h/$kpc&$R_{\Omega}/$kpc &\textbf{Refs.}\\
\hline
NGC 247&HSB&$2.7$&$3.5\pm0.6$&$[4]$\\
NGC 300&HSB&$1.8$&$1.9\pm0.3$&$[4]$\\
NGC 2403&HSB&$2.1$&$1.8\pm0.3$&$[4]$\\
NGC 3109&HSB&$1.35$&$1.6\pm0.3$&$[4]$\\
NGC 3769&HSB&$1.5$&$2.2\pm0.8$&$[1]$, $[2]$\\
NGC 3877&HSB&$2.4$&$2.6\pm0.4$&$[1]$, $[2]$\\
NGC 3893&HSB&$2.0$&$1.7\pm0.4$&$[1]$, $[2]$\\
NGC 3972&HSB&$1.6$&$2.5\pm0.5$&$[1]$, $[2]$\\
NGC 3917&LSB&$2.6$&$2.6\pm0.2$&$[1]$, $[2]$\\
NGC 4010&LSB&$2.9$&$3.8\pm0.8$&$[1]$, $[2]$\\
NGC 4085&HSB&$1.3$&$1.9\pm0.3$&$[1]$, $[2]$\\
NGC 4157&HSB&$2.2$&$2.4\pm0.5$&$[1]$, $[2]$\\
NGC 4183&LSB&$2.65$&$2.9\pm0.8$&$[1]$, $[2]$\\
NGC 4217&HSB&$2.4$&$2.4\pm0.5$&$[1]$, $[2]$\\
NGC 4236&HSB&$2.7$&$3\pm0.4$&$[4]$\\
NGC 4389&HSB&$1.2$&$1.9\pm0.9$&$[1]$, $[2]$\\
UGC 128&LSB&$6.8$&$7\pm1.5$&$[3]$, $[5]$\\
UGC 1230&LSB&$4.5$&$6\pm1.2$&$[3]$, $[5]$\\
UGC 5005&LSB&$4.4$&$4.5\pm0.6$&$[3]$, $[5]$\\
UGC 5750&LSB&$5.6$&$5.8\pm1$&$[3]$, $[5]$\\
UGC 5999&LSB&$4.3$&$5\pm 1$&$[3]$, $[5]$\\
UGC 6399&LSB&$2.0$&$2.1\pm0.4$&$[1]$, $[2]$\\
UGC 6446&LSB&$1.5$&$1.6\pm0.6$&$[1]$, $[2]$\\
UGC 6667&LSB&$2.45$&$2.3\pm0.2$&$[1]$, $[2]$\\
UGC 6917&LSB&$2.4$&$2.3\pm0.2$&$[1]$, $[2]$\\
UGC 6923&LSB&$1.1$&$1.8\pm0.2$&$[1]$, $[2]$\\
UGC 6930&LSB&$2.0$&$2.4\pm0.3$&$[1]$, $[2]$\\
UGC 6962&HSB&$1.4$&$1.6\pm0.5$&$[1]$, $[2]$\\
UGC 6983&LSB&$2.2$&$1.9\pm0.5$&$[1]$, $[2]$\\
F 561-1&LSB&$2.8$&$2.8\pm0.5$&$[3]$, $[5]$\\
F 563-V2&LSB&$2.1$&$1.8\pm0.3$&$[7]$, $[6]$\\
F 565-V2&LSB&$2.7$&$3.5\pm0.5$&$[3]$, $[5]$\\
F 568-3&LSB&$4.0$&$3.1\pm0.5$&$[3]$, $[6]$\\
F 568-V1&LSB&$3.2$&$2.7\pm0.4$&$[3]$, $[6]$\\
F 571-V1&LSB&$3.2$&$4\pm1$&$[3]$, $[5]$\\
\hline
\end{tabular}
\end{center}
\label{samplet}
\caption{The final sample. The references are coded as: $[1]$ Tully et al. (1996), $[2]$ Verheijen \& Sancisi (2001), $[3]$ de Blok et al. (1995), $[4]$ Kent (1987), $[5]$ de Blok \& McGaugh (1997), $[6]$ Swaters et al. (2000), $[7]$ McGaugh \& Bothun (1994).}
\end{table}
 
\begin{figure}
   \centering
\includegraphics[width=9cm, angle=0]{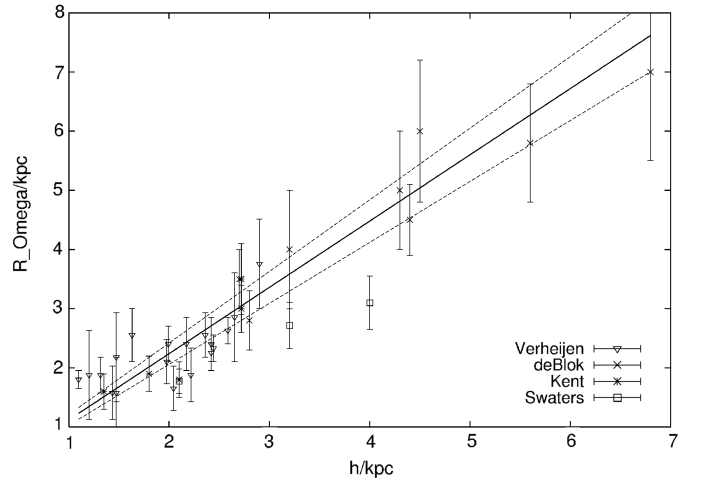}
      \caption{Correlation between
      the characteristic lengths $h$ and $R_{\Omega}$
      (see text for a description) for the
      sample of disk galaxies of Table~1. The dashed lines represent $3\sigma$ error on the best-fit value of the angular coefficient.}
         \label{corrfig}
   \end{figure}

When available (for example in the whole sample of galaxies belonging to the Ursa Major cluster), 
the reported exponential length is relative to the K-band photometry 
of the stellar disk, which is known to be the best representation of 
the actual stellar density distribution. In all the other cases we use 
the data as reported in the original source papers. We do not consider 
explicit error estimates for the exponential lengths $h$, since a 
uniform estimate would not be available for all the galaxies in the 
sample. The estimate for the radius $R_{\Omega}$ is calculated 
directly, following its definition Eq.~(\ref{romega}), from the 
observed rotation curve profile. The relevant error is estimated 
here by considering the errors in the velocity data-points and the 
process of interpolating the rotation curve between its measured points.
 
Figure \ref{corrfig} demonstrates that there is indeed a strong, tight
correlation between the two length scales considered.
The correlation found is
 
\begin{equation}
R_{\Omega}/h = 1.07 \pm 0.03~. 
\label{corr}
\end{equation}

\noindent The error on this measure is the standard statistical error, calculated by taking into account the uncertainties assigned to the values of $R_{\Omega}$.

Given the success of this test, it would be interesting to carry out a similar investigation on a better-defined, ``cleaner" sample of galaxies, so as to test how much of the scatter around the correlation is intrinsic or rather due to the quality of the available data. However, a discussion of this point would bring us away from the main objective of this paper.
  
The correlation coefficient for the entire sample is
$r_{h-R_{\Omega}} = 0.96$. It is thus interesting to note that the same
correlation applies to both HSB and LSB galaxies. The correlation
provides additional support to a conspiracy argument, because it
shows that even in the inner region of approximately linear growth
of the rotation curve the dark matter distribution must be in tune
with the distribution of the visible mass. Note that this is not
necessarily in favor of the maximum-disk hypothesis, but rather it
suggests that the relative distribution of visible and dark matter
in spiral galaxies is homologous among different objects. Such homology could also be at the basis of observed correlations such as the Tully-Fisher (1977) relation (van Albada \& Sancisi 1986).
 
In addition, this result will allow us to make a significant
(although preliminary) test of the models developed in this paper
on a simple idealized prototypical case that incorporates the
observed correlation, without the need for separate tests on
several observed objects.

\section{An isothermal halo embedding a zero-thickness disk}

\label{Sect:Problem}
 
We refer to standard cylindrical coordinates $(R, z, \theta)$, so
that $r = \sqrt{R^2+z^2}$ represents the spherical radius. We
assume that the rotation curve $V(R)$ of the disk galaxy, for
which we want to construct a mass model, is known (i.e., it has
been ``observed"); as a mathematical boundary condition, we assume
that the rotation curve remains flat at $V_{\infty}$ out to
infinite radii.
 
In order to make a fit to the observed rotation curve, we wish to
construct axisymmetric equilibrium models made of two mass
components: the visible disk and the dark halo. For simplicity, we
imagine the disk to be made only of stars, with known surface
brightness profile $\mu(R)$, and ignore the contribution of other
visible components. It will soon be clear that the contributions
from a gaseous disk and from a spheroidal bulge with given
structure could be easily incorporated, if desired. The visible
disk is described as a zero-thickness disk; in a follow-up paper,
we will address the study of a self-consistent finite-thickness
disk as a generalization of the present model.
 
The dark halo is taken to be characterized by a Maxwellian,
isothermal distribution function. The choice of the dark matter
distribution function is justified by an argument of physical
simplicity. With the minimum number of free parameters, it
guarantees the flatness of the outer parts of the rotation curve.
The procedure that we are going to describe can be easily adapted
to study dark halos characterized by other physically justified
distribution functions, if so desired (e.g., see Bertin, Saglia,
and Stiavelli (1992) and Bertin (2000)). In conclusion, for the dark matter halo we take the isotropic
distribution function:
 
\begin{equation}
f_{DM}(E) = {{\rho_{DM}^{0}}\over{\left(2\pi
\sigma^2\right)^{3/2}}}\exp{\left(-E/\sigma^2\right)}~,
\label{fiso}
\end{equation}
 
\noindent where $E = v^2/2 + \Phi_T$ is the specific energy of
dark matter particles and $\Phi_T$ is the \textit{total}
gravitational potential. In our model $\Phi_T$ is determined by
the dark matter halo \textit{and} by the disk:
 
\begin{equation}
\Phi_T = \Phi_{DM} + \Phi_D ~. \label{potential}
\end{equation}
 
\noindent The density distribution associated with the
distribution function (\ref{fiso}) is
 
\begin{equation}
\rho_{DM}(R,z)= \int d^3\vec{v}\ f_{DM} =
\rho_{DM}^{0}\exp{\left[-\Phi_T(R,z)/\sigma^2\right]}~,
\label{rhoiso}
\end{equation}
 
\noindent where we have made it explicit that, in general, it is
not spherically symmetric, because of the presence of the disk. At
large radii, the gravitational contribution of the disk becomes
vanishingly small, so that the total gravitational potential
eventually approaches the natural spherical symmetry of the
self-consistent isolated isothermal \textit{sphere}.
 
The dimensional free scales of the distribution function in
Eq.~(\ref{fiso}) are the central density $\rho_{DM}^{0}$ and
the (constant) velocity dispersion $\sigma$. We identify the
quantity $\rho_{DM}^{0}$ with the central dark matter density,
because we impose the condition
 
\begin{equation}
\Phi_T(0,0)=0~. \label{zeropot}
\end{equation}
 
\noindent In other words, we will only consider models that are
regular at the center.
 
The velocity dispersion $\sigma$ is uniquely fixed by the value
$V_{\infty}$ of the rotation curve at large radii, so that the
central dark matter density $\rho_{DM}^{0}$ is the only remaining
free dimensional scale for the adopted dark matter density
distribution. In fact, at large distances from the center, where 
the system is almost spherically symmetric, matching with the observed rotation curve requires
 
\begin{equation}
\Phi_T(R,z)\sim \Phi_T(r)\sim V_{\infty}^2
\ln\left({r\over{r_0}}\right)~, \label{asypot}
\end{equation}
 
\noindent which, inserted in the Poisson equation, gives
 
\begin{equation}
\label{asycon}
\left\{\begin{array}{lcl}
V_{\infty}^2 & = & 2\sigma^2\\
V_{\infty}^2 & = & 4\pi G \rho_{DM}^{0} r_0^2
\end{array}\right.~.
\end{equation}
 
\noindent We thus see that the scale length $r_0$ of the dark
matter density distribution appearing in Eq.~(\ref{asypot}) is not
a free length, because for assigned value of $V_{\infty}$ (i.e.,
of $\sigma$) it is determined by the value of the central dark
matter density, $\rho_{DM}^{0}$.
 
For our zero-thickness disk, we take the surface density to
be proportional to the observed surface brightness $\mu(R)$, i.e.:
 
\begin{equation}
\rho_D(R,z)=\left({M\over L}\right) \mu(R)\ \delta(z)~;
\label{rhodisk}
\end{equation}
 
\noindent here the function $\delta$ is the Dirac delta-function.
The ratio $M/L$ is the mass-to-light ratio of the stellar disk,
which we take to be constant; for assigned photometric profile
$\mu(R)$, it represents the only free dimensional scale of the
stellar density distribution.
 
The model is constructed by solving the complete Poisson equation
for the system, in which the matter densities are defined by the
distributions (\ref{rhoiso}) and (\ref{rhodisk}):
 
\begin{equation}
\nabla^2 \Phi_T(R,z) = 4\pi G \left[\rho_D(R,z) +
\rho_{DM}(R,z)\right]~. \label{dimpois}
\end{equation}
 
\noindent This is a nonlinear differential equation for the total
gravitational potential $\Phi_T$, which must be solved with the
appropriate boundary conditions: Eq.~(\ref{zeropot}), as the
condition at the center, and Eq.~(\ref{asypot}), for the
asymptotic behavior at large radii. The solution of the Poisson
Eq.~(\ref{dimpois}), which depends on the two free scales
$\rho_{DM}^{0}$ and $M/L$, will allow us to calculate the rotation
curve of the model $V_{mod}(R)=\sqrt{R(\partial \Phi_T/\partial R)|_{z=0}}$,
to be compared with the observed $V(R)$.
 
Note that this method is intrinsically different from the standard
parametric decompositions of the mass models of spiral galaxies. 
In standard parametric analyses, the contribution of the dark matter halo to the total
gravitational potential has a fixed shape (usually, spherically
symmetric), which is independent of the presence and gravitational
importance of the stellar disk and which can be changed only
through a set of parameters $a_i$, that is
$\Phi_{DM}=\Phi_{DM}(R,z;a_i)$. The present self-consistent models \textit{do not} have a fixed
parametric profile for the shape of the gravitational potential of the dark matter
component. Through Eq. (\ref{dimpois}), the dark matter gravitational potential, and thus its contribution to the total rotation curve, depends implicitly also on the properties of the stellar disk (its surface brightness $\mu(R)$ and its mass to light ratio), that is $\Phi_{DM}=\Phi_{DM}(R,z; \rho_{DM}^{0}; \mu(R), M/L)$.
 
In conclusion, for a given rotation curve $V(R)$ and surface
brightness of the stellar disk $\mu(R)$, the best-fit
self-consistent disk-halo decomposition will correspond to the
self-consistent model defined by the pair $(\rho_{DM}^{0}, {M/
L})$, together with the asymptotic velocity $V_{\infty}$, that gives the best fit to the observed rotation curve.

\subsection{The mathematical problem in dimensionless form}

As units of length and of specific energy we will consider the scale length of the stellar disk $h$ and
the square of the velocity dispersion $\sigma^2=V_{\infty}^2/2$,
respectively, so that we introduce the dimensionless coordinates

\begin{equation}
\xi\equiv R/h~,\   \zeta\equiv z/h~,\ \eta\equiv r/h
\label{lengths}
\end{equation}
 
\noindent and the dimensionless gravitational potential
 
\begin{equation}
\psi_T=\psi_{DM}+\psi_{D}\equiv -\Phi_T/(V_{\infty}^2/2)~.
\end{equation}
 
\noindent We separate the Poisson Eq.~(\ref{dimpois}) in the
two Poisson equations for the two different mass components of the
system. The dimensionless equation for the stellar disk reads
 
\begin{equation}
\left({1\over\xi}{\partial\over{\partial\xi}}\xi{\partial\over{\partial\xi}}
+{{\partial^2}\over{\partial\zeta^2}}\right)\psi_{D}=-{\beta\over
2}\  \hat\Sigma(\xi)\delta(\zeta)~. \label{poisdisk}
\end{equation}
 
\noindent Here we have introduced the normalized surface density,
defined by $\hat\Sigma(R)\equiv{{\Sigma(R)}/{\Sigma(0)}}$, and the
parameter $\beta$
 
\begin{equation}
\beta\equiv{{8\pi G\Sigma(0)h}\over{V^2_{\infty}}} ={M\over L}\
{{8\pi G\mu(0)h}\over{V^2_{\infty}}}~, \label{beta}
\end{equation}
 
\noindent which is the dimensionless version of the weight of the
stellar disk as measured by the mass-to-light ratio $M/L$.
 
To solve Eq.~(\ref{poisdisk}), we may use the standard
technique of Hankel transforms (see Toomre (1963)). In this way
the potential associated with any surface density profile $(M/L)\mu(R)$
can be calculated exactly. Consistent with the condition
(\ref{zeropot}), we also impose that $\psi_D(0,0)=0$. In particular,
for purely exponential disks, $\hat\Sigma=\exp(-\xi)$, the
associated gravitational potential is given by
 
\begin{equation}
\psi_D(\xi, \zeta)=-{\beta\over 2}\left\{ \int_{0}^{\infty} d(hk)
{{\exp\left(-hk~\left|\zeta\right|\right)J_{0}(hk~\xi)}\over{\left[1+\left(hk~\xi\right)^2\right]^{3/2}}} -1\right\}~,
\end{equation}
 
\noindent where $J_0$ follows the standard notation of Bessel
functions.
 
In turn, the Poisson equation for the dark matter halo can be
written as
 
\begin{equation}\label{poishalo}
\left({1\over\xi}{\partial\over{\partial\xi}}\xi{\partial\over{\partial\xi}}
+{{\partial^2}\over{\partial\zeta^2}}\right)\psi_{DM}=-\alpha
\exp{\left[\psi_{DM}(\xi,\zeta)+\psi_{D}(\xi,\zeta)\right]}~,
\end{equation}
 
\noindent where the parameter $\alpha$ can be seen as the
dimensionless expression of the central dark matter density scale:
 
\begin{equation}
\alpha\equiv {{8\pi G h^2\rho_{DM}^0}\over{V^2_{\infty}}} = 2{{h^2}\over {r_0^2}} ~.
\label{alpha}
\end{equation}

\noindent The parameter $\alpha$ measures the halo concentration. With an eye to the central region of the system, a measure of the halo mass relative to the the mass of the disk is given by the parameter $\alpha/\beta = \rho_{DM}^{0} h / \Sigma_0$ (see also Eq.~(\ref {dimpois})). Thus the dimensional pair $(\rho_{DM}^{0}, {M/ L})$ has its dimensionless counterpart in $(\alpha, \beta)$. In the $(\alpha, \beta)$ plane, the straight lines defined by $\beta \propto \alpha$ can be interpreted as lines of approximately given central density of the halo with respect to the mass of the disk; steep lines are those characterizing relatively light halos. A better defined measure of the weight of the halo will be given in the next Section.
 
The Poisson Eq.~(\ref{poishalo}) for the dark matter halo has
to be solved numerically (see Appendix A). Here the
gravitational potential of the disk, calculated separately as
described above, acts as an external potential that breaks the
spherical symmetry and shapes the dark matter density
distribution. The relevant boundary conditions can be obtained
from the boundary conditions imposed on the total gravitational
potential:
 
\begin{equation}\left\{
\begin{array}{lcl}
\psi_{DM}(0,0)&=&0\\
\psi_{DM}(\eta)&\sim&-2\ln\left(\eta \sqrt{\alpha \over 2}\right)-\psi_{D}^{\infty}\ ,\  \eta\gg 1
\end{array}\right.~,
\label{boundary}
\end{equation}
 
\noindent where the symbol $\psi_{D}^{\infty}$ indicates the
constant value approached (with spherical symmetry) by the 
dimensionless gravitational potential of the disk at large 
distances from the center of the system. 

\section{Properties of self-consistent spheroidal isothermal halos}\label{Sect:Properties}

The structure of the dark matter halo reflects the structure of
the total gravitational potential, as prescribed by
Eq.~(\ref{rhoiso}).

On the one hand, there is an artificial cusp in the vertical dark
matter density distribution on the $z=0$ plane, which results from
the singular (zero-thickness) density distribution assumed for the
stellar disk. The detailed structure of the cusp can be obtained
from the Jeans equation:

\begin{equation}
{{\sigma^2}\over{\rho_{DM}(R,0)}}\left[{{\partial}\over{\partial
z}}\rho_{DM}(R,0^{+})-{{\partial}\over{\partial
z}}\rho_{DM}(R,0^{-})\right]
=\left.{{\partial \Phi_{D}}\over{\partial
z}}\right|^{(R,0^{+})}_{(R,0^{-})}~. \label{cuspide}
\end{equation}

\noindent This feature has no physical relevance and will be
naturally eliminated in the finite-thickness disk models that we
plan to present in a separate paper.

\begin{figure}
   \centering
\includegraphics[width=8.7cm, angle=0]{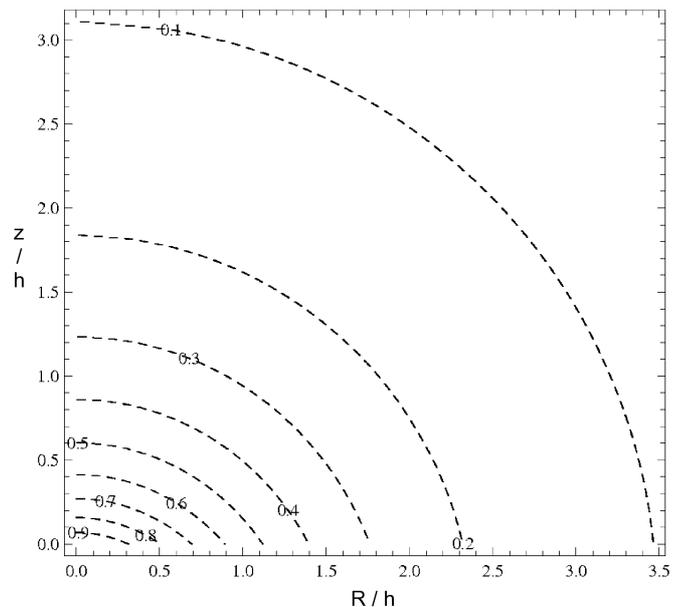}
      \caption{Contours of the dark halo density distribution (normalized to its central value) for the model defined by Eq.~(\ref{bf}). Note that these contours coincide with those of the total gravitational potential.}
         \label{contours}
   \end{figure}

On the other hand, the halo is characterized by significant
flattening. This latter feature is genuine and will not disappear
in the regular finite-thickness case. Figure \ref{contours} shows
the contours of the total gravitational potential, which coincide
with those of the dark matter density, for the model defined by
$(\alpha, \beta) = (1.9, 3.4)$, which, as we will see in the
following, is the self-consistent model representative of the
correlation in Eq. (\ref{corr}).

\subsection{The shape of isothermal halos}\label{shape}

\begin{figure}
   \centering
\includegraphics[width=8.7cm, angle=0]{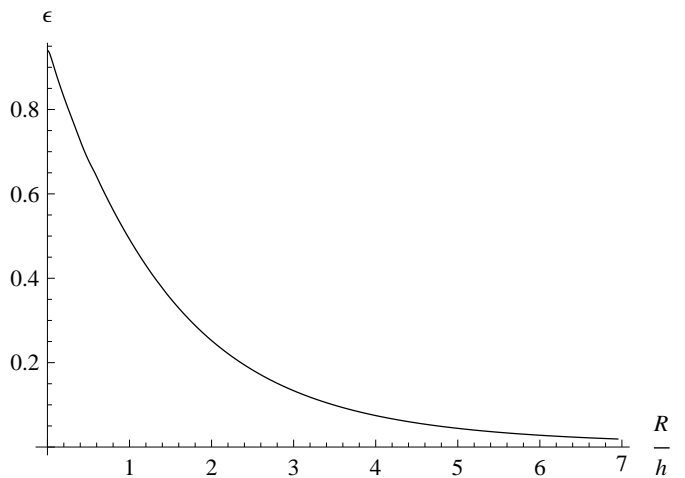}
      \caption{Ellipticity profile of the dark matter halo illustrated in Fig.~\ref{contours}, as defined by Eq.~(\ref{ells}).}
         \label{flattening}
\end{figure}

\begin{figure}
   \centering
\includegraphics[width=8.7cm, angle=0]{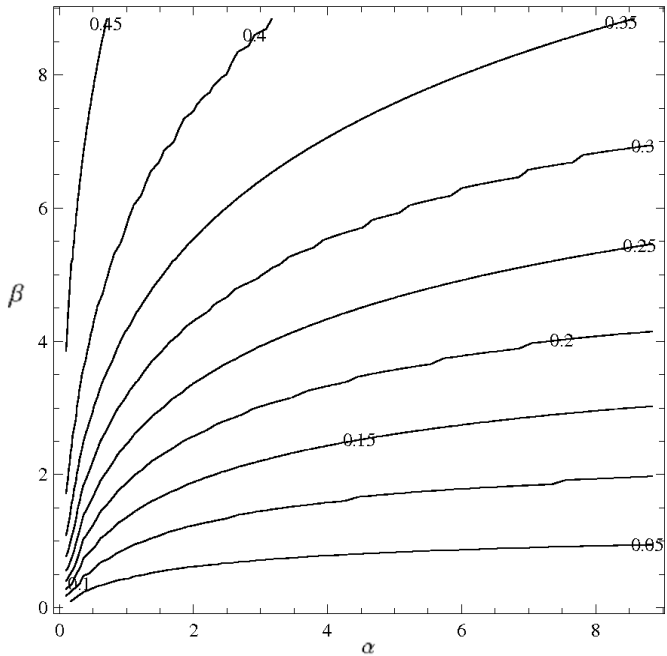}
      \caption{The ellipticity at $2.2$ exponential lengths $\varepsilon(2.2)$ in the $(\alpha, \beta)$ plane.}
         \label{ellplane}
\end{figure}

As naturally expected, the system tends to be spherically
symmetric in the outer parts (where the gravitational importance
of the disk rapidly vanishes), but is highly oblate in the central
regions. The flattening of the dark halo can be quantified in
terms of the ellipticity $\varepsilon$, defined as

\begin{equation}
 \varepsilon(\xi)=1-{{v(\xi)}\over {\xi}}~,
 \label{ells}
 \end{equation}

\noindent where $\xi$ and $v(\xi)$ represent respectively the
intersections with the $\zeta=0$ and $\xi=0$ planes of a given
contour of the total gravitational potential. The profile of the
ellipticity $\varepsilon$ as a function of radius is illustrated
in Fig.~\ref{flattening}. The high flattening inside one exponential 
length may be exaggerated because of the zero-thickness of the disk.
Figure~\ref{ellplane} shows the ellipticity reached by the self-consistent models at $2.2$ exponential lengths in the $(\alpha, \beta)$ plane.

It would be interesting to compare the flattening obtained from our self-consistent models with the empirical determinations based on a variety of dynamical arguments, ranging from the dynamics of warped disks and polar rings to the flaring of gas layers (see the discussion in the paper by Olling \& Merrifield 2000; see also the series of papers started by O'Brien et al. 2010). It would also be interesting to compare the flattening of our self-consistent models with the indications that are expected to be gathered in the near future from weak and strong gravitational lenses (for example, see Trott et al. 2010); here, the natural comparison to be made is in terms of the relevant flattening profile of the projected density distribution.

\subsection{The flatness of self-consistent rotation curves}\label{flat}

A significant feature of the self-consistent models presented in
this paper is the unexpected degree of flatness that characterizes
their rotation curves. In principle, self-consistent rotation
curves are designed to be flat only as an asymptotic condition for
$R \gg h$, that is where the gravitational influence of the
stellar disk becomes negligible. In the intermediate radial range
the shape of the rotation curve is instead determined by the
mutual interactions between halo and visible disk and thus cannot be 
ascribed naively to the Maxwellian form of the halo distribution function 
(especially because in the intermediate radial range an isothermal sphere is known 
to be associated with a rotation curve exhibiting a characteristic oscillation
pattern).

To better demonstrate this point and to clarify the origin of the
observed flatness, the self-consistent rotation curves will be
compared with those of two different non-self-consistent models:
(i) the parametric models, defined by Eqs.~(\ref{vd},\ref{vdm});
(ii) the models, called for short \textit{semi--parametric}, for
which the field of the disk, with associated rotation curve
defined by Eq.~(\ref{vd}), is superposed (non-self-consistently)
to the field of a \textit{spherical} isothermal halo, i.e. a halo
described by the density distribution in Eq.~(\ref{rhoiso}), in
which the \textit{total} potential $\Phi_T(R,z)$ is replaced by
the dark matter potential $\Phi_{DM}(r)$. Therefore, any
difference in the rotation curves between semi-parametric models
and self-consistent models must be ascribed to the effects of
self-consistency alone.

With the aim of providing a systematic and quantitative analysis,
we start by defining as intermediate radial range the range $2.2h
< R < 7h$, i.e. $2.2 < \xi < 7$. Here the lower limit corresponds
to the location at which the rotation curve of an isolated
exponential disk reaches its maximum, while the upper limit
represents a reasonable choice of a location outside the bright
optical disk. The quantity $\hat{V}_{\infty}$ will denote the
average value of the rotation curve $V_{mod}$ in the \textit{outer
parts}, $4 < \xi < 7$. We then introduce the function

\begin{equation}
\Theta(\alpha, \beta)=
\int_{2.2}^{7} \left|{1\over{ \hat{V}_{\infty} } } {{\partial
V_{mod}}\over{\partial \xi}}\right| d\xi~.\label{flatness}
\end{equation}

\begin{figure}
  \centering
\includegraphics[width=8.7cm, angle=0]{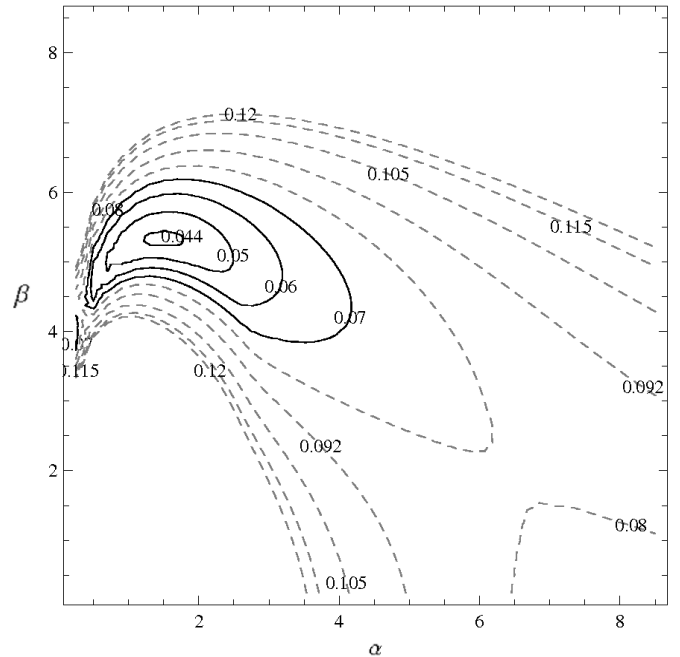}
     \caption{Contours of the function $\Theta$ (Eq.~(\ref{flatness})) for
     self-consistent models.}
        \label{selfflat}
  \end{figure}

\begin{figure}
  \centering
\includegraphics[width=8.7cm, angle=0]{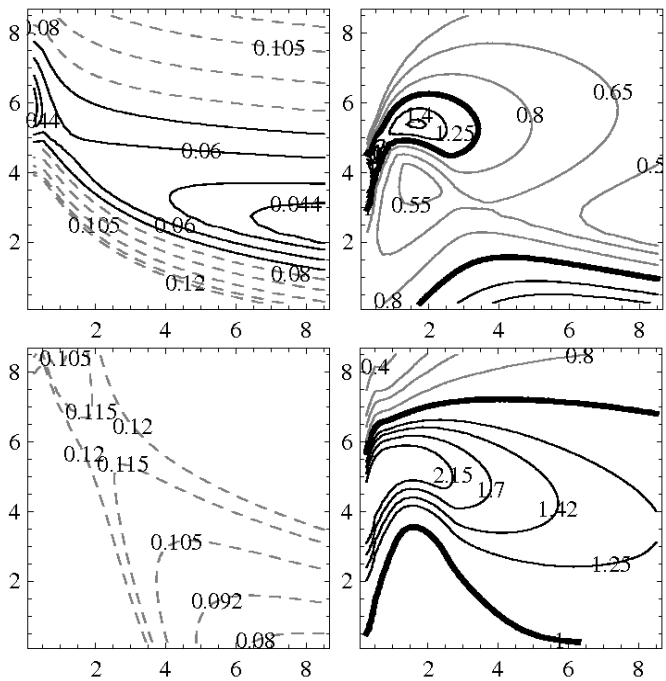}
     \caption{Top-left: Contours of the function $\Theta$ for parametric mass models.
     Top-right: Contours of the ratio between the function $\Theta$ for parametric
     models and the same function for self-consistent models. Bottom-left:
     Contours of the function $\Theta$ for semi-parametric models,
     with isothermal \textit{spherical}
     halo. Bottom-right: Contours of the ratio between the
    function $\Theta$ for semi-parametric models and the same function for
    self-consistent models. In the right frames, the thick contour represents unity.
    The axes in the
    four panels are $(\alpha, \beta)$, as in Fig.~\ref{selfflat}.}
        \label{flatpanels}
  \end{figure}

\noindent The function $\Theta$ is constructed so as to measure
the flatness of the rotation curve $V_{mod}$ in the intermediate
radial range. The $\Theta$-contours calculated from the
self-consistent models are displayed in Fig.~\ref{selfflat}. In
Fig.~\ref{flatpanels} the two left panels show the contours of the
function $\Theta$ obtained from the parametric models (upper
frame) and the semi-parametric models (lower frame), for the same
region of parameter space. A set of representative contours has been chosen
arbitrarily, with the aim of describing the properties of the
three different cases.

With respect to the range of values of the function $\Theta$, the
parametric models and the semi-parametric models differ
significantly from each other, because the semi-parametric models
exhibit rotation curves that are less flat over the entire region
of parameter space. This behavior is due to the fact that regular
isothermal spheres are characterized by a well-known oscillation
pattern in their density distribution and in the associated
rotation curve, so that in the inner region the rotation curve
overshoots its asymptotic value (see also Fig.~\ref{panels1}).
This property is removed \textit{a priori} in the parametric
profile of Eq.~(\ref{vdm}), often adopted in this context. In
contrast, self-consistent models show a degree of flatness
comparable to that of parametric models. The right panels of
Fig.~\ref{flatpanels} illustrate (upper frame) the contours of the
ratio between the function $\Theta$ for parametric models and the
same function for self-consistent models and (lower frame) the
contours of the ratio between the function $\Theta$ for
semi-parametric models and the same function for self-consistent
models. It is then clear that, in a relatively wide region of
parameter space, where the contour levels exceed unity in the
lower right frame of Fig.~\ref{flatpanels}, self-consistent
rotation curves are more flat than the corresponding
semi-parametric rotation curves; as stated previously, this
\textit{flattening effect} can only be due to self-consistency
(this point will be addressed further in Sect.~\ref{shaperc}). In
addition, there is a region of parameter space in which
self-consistent rotation curves are more flat than parametric
rotation curves (where the contour levels exceed unity in the
upper right frame of Fig.~\ref{flatpanels}). This latter
comparison has no special physical meaning, but it confirms the
high degree of flatness of self-consistent rotation curves (which
is also evident from Fig.~\ref{panels}).

Another important point of comparison is the shape of the contours
of the function $\Theta$ for the different cases. While the
self-consistent models are flat in a well-defined region of
parameter space (e.g., see the distinction between solid and dashed
contours in Fig.~\ref{selfflat}), the parametric models are
uniformly flat over an entire diagonal strip of parameter space
(upper left frame of Fig.~\ref{flatpanels}). In Sect.~\ref{decomp}
we will see that this strip is remarkably similar to the disk-halo
degeneracy strip.

\subsection{The inner gradient of the self-consistent rotation curves}\label{romegasect}

\begin{figure}
   \centering
\includegraphics[width=8.7cm]{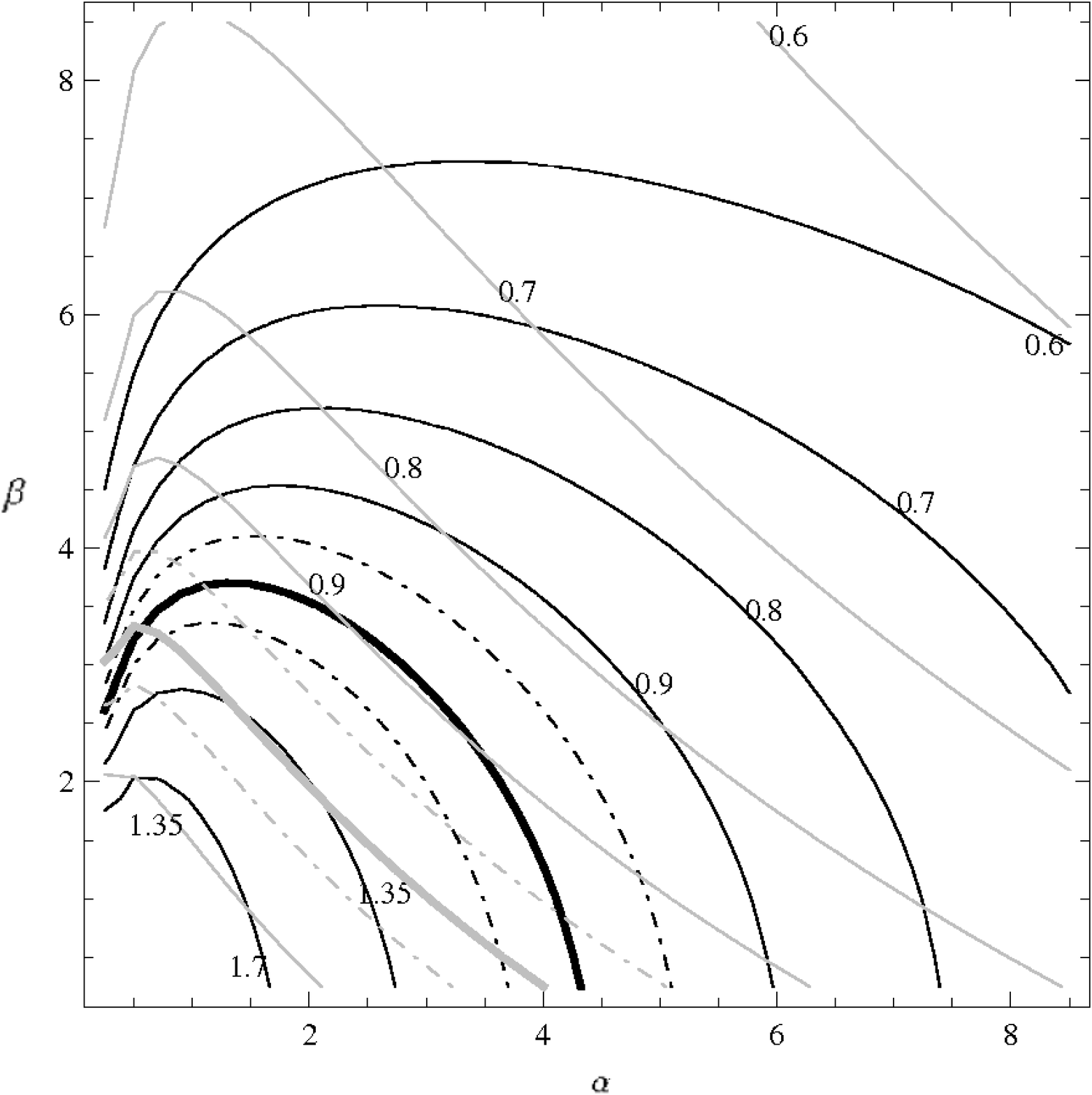}
\caption{Contours of the quantity $R_{\Omega}/h$, defined in
Eq.~(\ref{romega1}). The gray contours, exhibiting a diagonal
trend, are associated with the parametric models defined by
Eqs.~(\ref{vd}) and (\ref{vdm}), while the black contours are
associated with the self-consistent models. In both cases the
thick line represents the empirical value found in
Eq.~(\ref{corr}), and the dashed lines represent the edges of the
interval defined by a $3\sigma$ deviation.} \label{figromega}
 \end{figure}

After characterizing the properties of the self-consistent models
at intermediate radii, in this subsection we focus on the
properties of the inner regions. We start by defining the radius
$R_{\Omega}$ from the relation

\begin{equation}
V(R_{\Omega})={2\over 3}\hat{V}_{\infty}~, \label{romega1}
\end{equation}

\noindent with $\hat{V}_{\infty}$ introduced in Sect.~\ref{flat}
Figure~\ref{figromega} shows the contours of the quantity
$R_{\Omega}/h$ for self-consistent models (black contours) and
parametric models (gray contours). The thick contours in
Fig.~\ref{figromega} represent the contours corresponding to the
empirical correlation found in Eq.~(\ref{corr}).

The general shape of the contours is significantly different for
the two models. While self-consistent models show roundish
contours, parametric models exhibit a clear diagonal trend, which
reinforces the expectation that a degeneracy problem will arise in
the corresponding disk-halo decompositions, as noted in the
previous subsection. 

Finally, in the case of self-consistent models, the general shape
of the $R_{\Omega}/h$ contours is similar to that of the contours
of the function $\Theta$. The $R_{\Omega}/h$ value corresponding
to maximum-flatness is slightly below the one suggested
empirically by the analysis of Sect.~\ref{empiric}, that is it
corresponds to a slightly faster growing rotation curve (but see
also the case of NGC 3198, described in Sect.~\ref{3198}).

\subsection{Systematic properties of the self-consistent solutions in the available parameter space}\label{shaperc}

The aim of this subsection is to provide a description of the
properties of self-consistent rotation curves in the relevant
parameter space.

Figure~\ref{panels} shows the rotation curves and the related
disk-halo decompositions for an atlas of self-consistent models,
in the region of parameter space of astrophysical interest (see
also Sect.~\ref{decomp}). In each frame the disk contribution to
the rotation curve is dash-dotted, the contribution of the
self-consistent halo is the solid black curve and the resulting
rotation curve is indicated by a thick curve. Each frame records
the pair $(M_{3.5}, M_{7.0})$. The quantities $M_{3.5}$ and
$M_{7.0}$ indicate the ratio of the halo mass to the disk mass
inside the spheres with radii of $3.5$ and $7$ exponential lengths, respectively.

As already noted in Sect.~\ref{flat}, in general the rotation
curves are significantly flat and lack specific features. The
flattening effect described in Sect.~\ref{flat} shows up clearly
in Fig.~\ref{panels}, in a sort of \textit{suppression effect}.
Here the effects on the self-consistent halo caused by the
presence of the stellar disk can be followed by looking at each
individual column, that is by fixing the value of the
dimensionless central density of the halo $\alpha$ and by letting
the weight of the disk $\beta$ increase. The parameter $\alpha$
fixes the innermost gradient of the dark halo contribution to the
rotation curve (which is proportional to $\sqrt{G \rho_{DM}^0}$).
However, in general self-consistency cannot support a fast growing
contribution from the dark halo in the presence of an important
stellar disk, and then it ``softens" it near $1\div 2$ 
exponential lengths. At small values of $\beta$
self-consistency suppresses the bump in the halo contribution to
the rotation curve characteristic of the isothermal sphere (e.g.,
compare the $(\alpha=9, \beta=0.5)$ frame to the $(\alpha=9,
\beta=2)$ frame). For heavier disks (large values of $\beta$),
self-consistency suppresses the halo contribution at intermediate
radii (e.g., compare the $(\alpha=0.5, \beta=3.5)$ frame to frames
corresponding to higher values of $\beta$). For high values of
both $\alpha$ and $\beta$ (see for example $(\alpha=9, \beta=8)$)
self-consistency makes the halo contribution fairly constant, but
with a value which is significantly below 100 \%, which will be
reached eventually, at much larger radii.

This \textit{suppression effect} has three main consequences on the general
shape of self-consistent rotation curves: it flattens them
globally (with respect to the rotation curves of the
semi-parametric models; see Sect.~\ref{flat}); it removes the
oscillations characteristic of the isolated isothermal sphere
(provided the stellar disk is not too light); it makes the
signature of the presence of an important luminous disk more
evident, by lowering the halo contribution.

Similar general conclusions can be made by inspection of
Fig.~\ref{panels1}. Total rotation curves of self-consistent
models (thick gray lines), parametric models (black solid lines),
and semi-parametric models (dashed lines) are shown here in
comparison, in a general format and layout similar to those of
Fig.~\ref{panels}. In each frame, the stellar disk contribution is
given by the dash-dotted line, while the constant line
$V_{mod}=V_{\infty}$ is drawn to make the comparison easier. What
clearly catches the eye is the remarkable uniformity of the 
parametric rotation curves, when compared to those of 
the semi-parametric models and of the self-consistent 
models. From a physical point of view, this uniformity is not justified, because it confirms that the parameterization used flattens out arbitrarily
the properties of the physical isothermal solutions that 
it is meant to represent approximately.

\subsection{Conspiracy and degeneracy}

Figure~\ref{figromega} demonstrates that the available parameter
space cannot accommodate values of $R_{\Omega}/h$ significantly
different from unity, especially if we also require that the rotation curve be flat at
intermediate radii. If galaxies were found with $R_ {\Omega}$
too small or too large with respect to $h$, there would be no way
for the picture of disks embedded in quasi-isothermal halos to
hold and then one should be led to look for dark matter
distributions significantly different from those characteristic of
isothermal halos. This point will be further addressed in Appendix
B, where the difficulty of fitting such ``anomalous" rotation
curve is demonstrated. Fortunately, the observed correlation
Eq.~(\ref{corr}) shows that the picture of a quasi-isothermal halo
is indeed viable.

\begin{landscape}

\begin{figure}
   \centering
\includegraphics[width=1.2\textwidth]{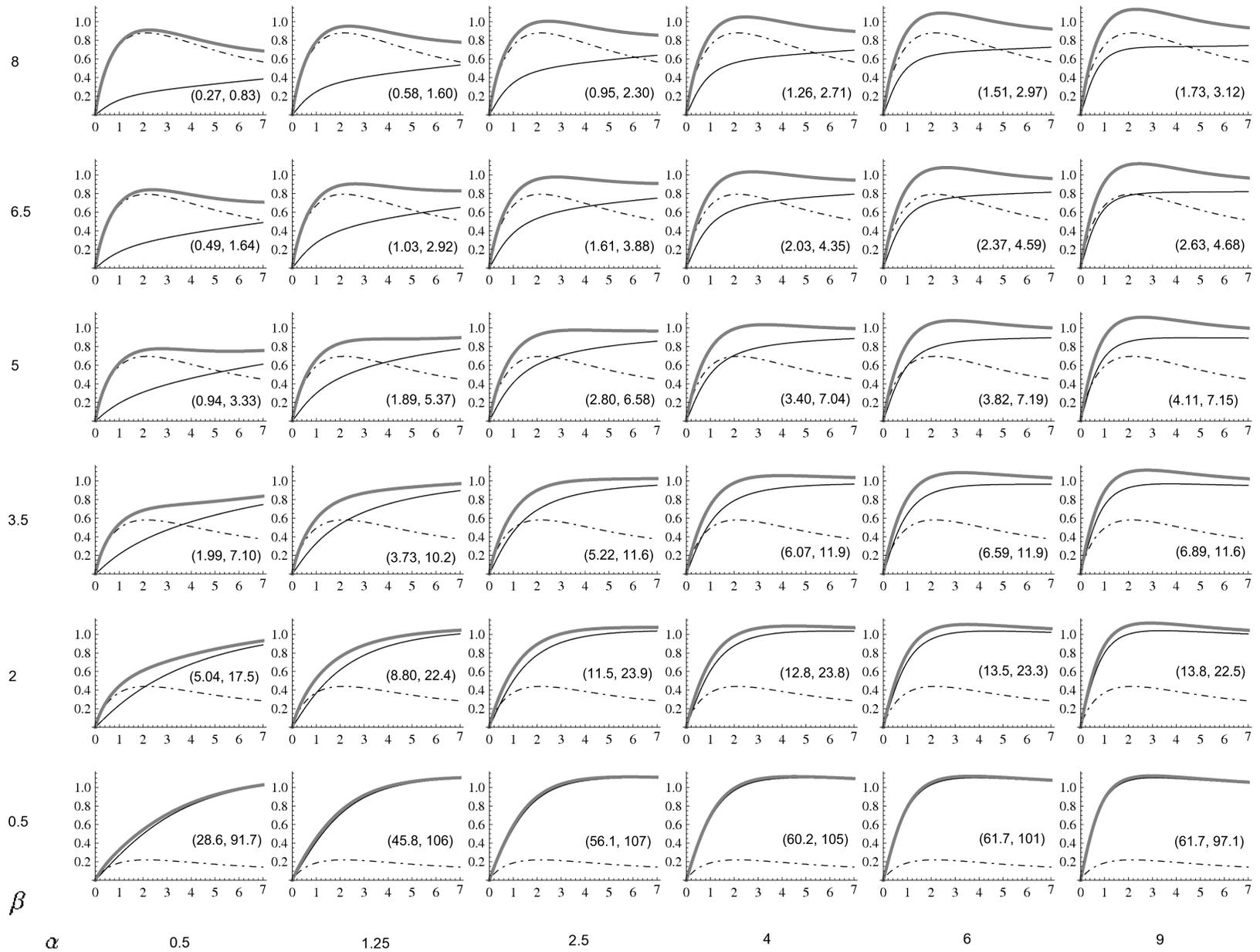}
\caption{Rotation curves in parameter space. The format and coding
are the same as in Fig.~\ref{bfrc}. The functions drawn are
$V/V_{\infty}$ versus $R/h$. For a description of the pairs
$(M_{3.5}, M_{7.0})$ recorded on the lower-right part of each
frame, see Sect.~\ref{shaperc}.} \label{panels}
   \end{figure}

\end{landscape}

\begin{landscape}

\begin{figure}
   \centering
\includegraphics[width=1.2\textwidth]{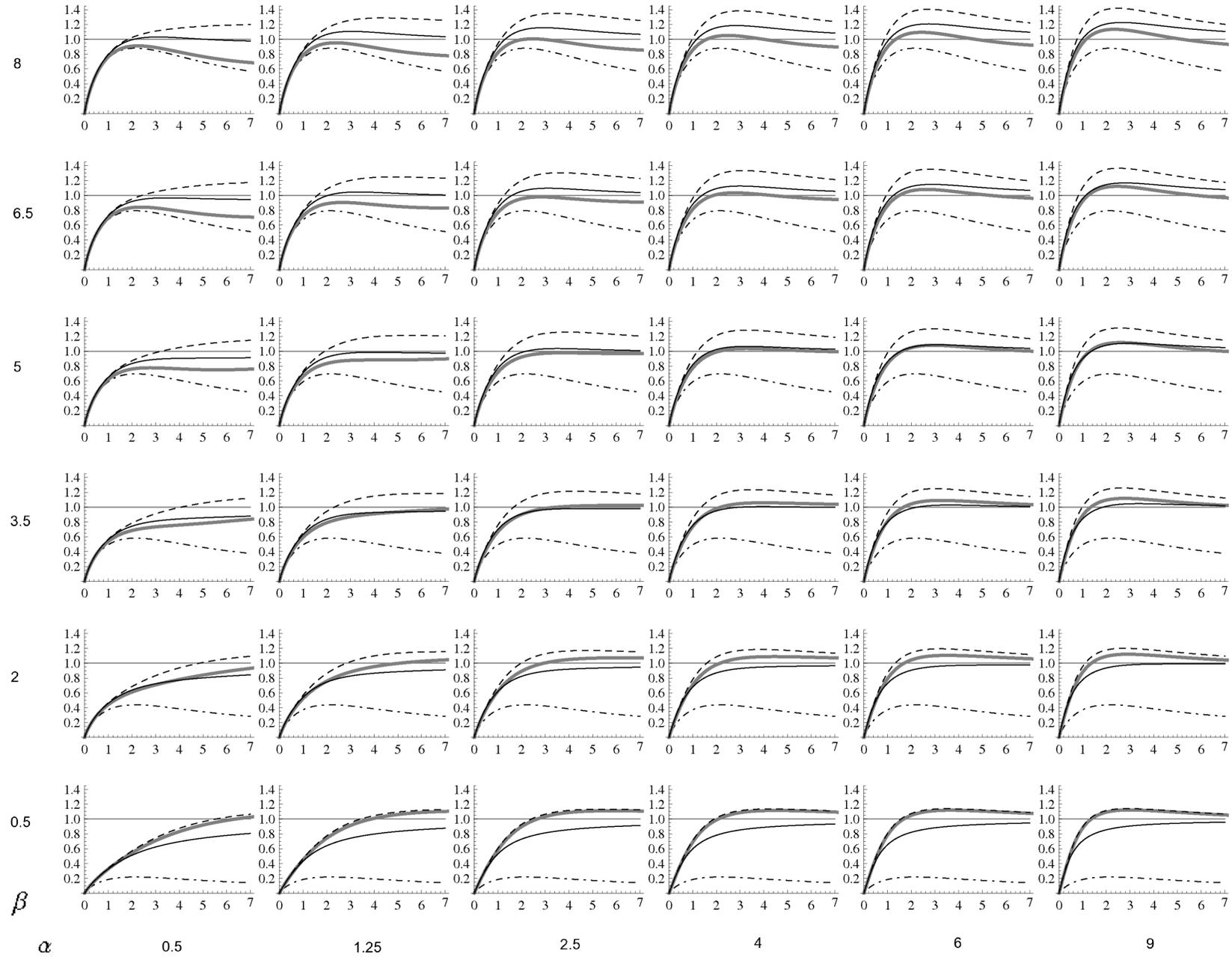}
\caption{Rotation curves in comparison. The functions drawn are
$V/V_{\infty}$ versus $R/h$. Thick gray lines are self-consistent
rotation curves, black lines are parametric rotation curves,
dashed lines are semi-parametric rotation curves (based on
\textit{spherical} isothermal halos); dash-dotted curves represent
the disk contribution.} \label{panels1}
\end{figure}

\end{landscape}

Figure~\ref{panels} confirms that self-consistent models lead to
rotation curves that are realistic, in the sense that they are
smooth and featureless for a wide range of disk-to-halo mass
ratios; note that the maximum disk solution associated with 
the correlation~(\ref{corr}) (see Sect.~\ref{degremoval} for description)
 is expected to occur at $\beta \approx 5$ (cf. Eq.~(\ref{maxdisk})), approximately in 
the middle of the parameter range discussed and explored in this 
Section. An inspection of Fig.~\ref{panels1} shows that parametric
models are also characterized by this property, actually in an
even more prominent way. In a sense, these findings demonstrate
that we are not facing an issue of conspiracy or, at least, that
the conspiracy is not as dramatic as often stated. In fact,
differently from what is naturally expected and is often reported,
no fine tuning is actually required in this problem, because for
very different values of the halo-to-disk mass ratio the rotation
curves resulting from the combination of a disk with a
quasi-isothermal halo, turn out to be
(surprisingly) smooth and flat. In this respect, the correlation represented by Eq. (4) implies a much tighter coupling between dark and luminous components
in the $(\alpha, \beta)$ plane.

In the context of parametric models, these conclusions appear to
be quite artificial. In fact, the use of the simple model
associated with Eq.~(\ref{vdm}) removes \textit{a priori}
significant features that are characteristic of the regular
isothermal sphere. Such non-natural uniformity in parameter space ends up in
generating the problem of disk-halo degeneracy. In fact, the
diagonal trend of the parametric models exhibited in the
upper-left frame of Fig.~\ref{flatpanels} and in
Fig.~\ref{figromega} reinforces the diagonal trend exhibited by
the same models in Fig.~\ref{parametric}, thus anticipating a
problem of degeneracy of solutions along diagonal lines in
parameter space.

The self-consistent models, instead, have a distinctly different
behavior (see Fig.~\ref{selfflat} and Fig.~\ref{figromega}); for them, maximal flatness
occurs in the vicinity of a single point of parameter space. We
are then brought to conclude that degeneracy may be broken by
self-consistency. This last point is confirmed and better
quantified in the following separate Section.

\section{Self-consistent disk-halo decomposition: the removal of the disk-halo degeneracy}\label{decomp}

\subsection{A preliminary test on a fiducial rotation curve}\label{degremoval}

As anticipated briefly in Sect.~\ref{empiric}, the empirical
correlation (\ref{corr}) found between the two relevant
scale lengths $h$ and $R_{\Omega}$ allows us to make a preliminary test
by focusing on a single, representative idealized case. Therefore,
a discussion of specific observed objects, such as NGC 3198, is
postponed and will be carried out separately (see following Sect.~\ref{3198})
 
As a typical reference case, we consider a galaxy characterized by
a purely exponential stellar disk, $\hat{\Sigma}(R) \equiv
\exp(-\xi)$, and by the following rotation curve
 
\begin{equation}
V(\xi, \tau)/V_{\infty}=1-\exp(-\xi/\tau)~. 
\label{paramprof}
\end{equation}
 
\noindent If we take
 
\begin{equation}
\tau\approx 1~,\label{taucorr}
\end{equation}
 
\noindent the correlation found in Eq.~(\ref{corr}) is reproduced.
 
The simple analytical form of the adopted rotation curve
(\ref{paramprof}) allows us to calculate explicitly the value of
the ``observed" parameter $\alpha_{obs}$ defined in
Eq.~(\ref{alphaobs}):
 
\begin{equation}
\alpha_{obs}=\alpha_{obs}(\tau)=\exp(2\gamma)/2\tau^2~,
\end{equation}
 
\noindent where $\gamma \approx 0.577$ denotes the Euler gamma
constant, thus $\alpha_{obs}(\tau = 1) \approx 1.59$; the following discussion will be based on the use of this value of $\alpha_{obs}$.
 
Note also that, for a given value of $\tau$, the rotation curve
(\ref{paramprof}) will admit a well-defined \textit{maximum-disk}
decomposition. In practice, before carrying out any detailed
decomposition, we may introduce, as a simple definition of
maximum-disk, the disk characterized by the value of the
dimensionless weight $\beta_{max}(\tau)$ that gives, without any dark
matter contribution, the best fit to the inner rotation curve in
the interval $\left[0,R_{\Omega}\right]$. Note that this maximum-disk 
decomposition is designed in such a way to recover the central gradient 
of the rotation curve. Sometimes a different definition is used, by referring to the
stellar disk which reaches the asymptotic rotation velocity in correspondence
of the maximum of its rotation curve. This definition gives higher values of $\beta$
($\beta_{max}\approx 10$ for an exponential disk), but it is completely independent
 of the properties of the central part of the rotation curve.
 
For $\tau$ given by equation (\ref{taucorr}), we find
 
\begin{equation}
\beta_{max}(1)\approx 5\ ; 
\label{maxdisk}
\end{equation}
 
\noindent this value of $\beta$ identifies an important reference value to
which the results obtained from the self-consistent decomposition
will thus be compared. We recall that $\beta=5$ is approximately in the middle
of the parameter range discussed and explored in Sect.~\ref{shaperc}. 

The properties of the self-consistent models constructed in the
present paper are best illustrated by describing their ability to
fit the just described representative idealized case in comparison
to a fit performed by more standard parametric analyses. For both methods the
goodness of a disk-halo decomposition, defined by a pair
$(\alpha,\beta)$, is quantified by the function
 
\begin{equation}
\Xi(\alpha, \beta, \tau)=
{1\over{V^2_{\infty}}}\int_{0}^{7}\left[V_{mod}(\xi)-V(\xi,
\tau)\right]^2\ d\xi~, \label{deviat}
\end{equation}
 
\noindent that is the integrated squared residuals between the
``observed" rotation curve and the rotation curve calculated from
the model $V_{mod} = \sqrt{{V}^2_D + V_{DM}^2}$. The cut of the integration at seven exponential 
lengths is a reasonable choice, consistent with the analysis of Sect.~\ref{flat} . 
In this preliminary test we do not treat the asymptotic velocity as an 
additional free parameter of the model. In other words, we just analyze
the deviations defined by Eq. (29) in the $(\alpha, \beta)$ plane, at fixed  $V_{\infty}$.  
The asymptotic velocity will be kept as a free parameter in Sect. 5.2,
for the case of NGC 3198.

\begin{figure}
   \centering
\includegraphics[width=8.7cm, angle=0]{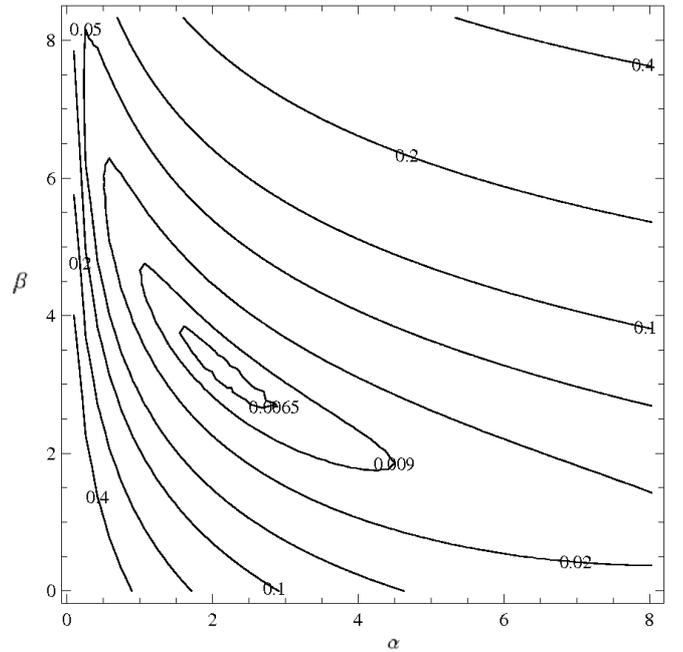}
      \caption{Contours of the function $\Xi$ for the
       case of the parametric decomposition based on Eqs.~(\ref{vd}, \ref{vdm}). As in similar plots (Figs.~\ref{nparametric}, \ref{ph}, \ref{nnph}, \ref{pl}, \ref{nnpl}), the displayed contours are chosen arbitrarily to illustrate the properties of the $\Xi$ function.}
         \label{parametric}
   \end{figure}
\begin{figure}
   \centering
\includegraphics[width=8.7cm, angle=0]{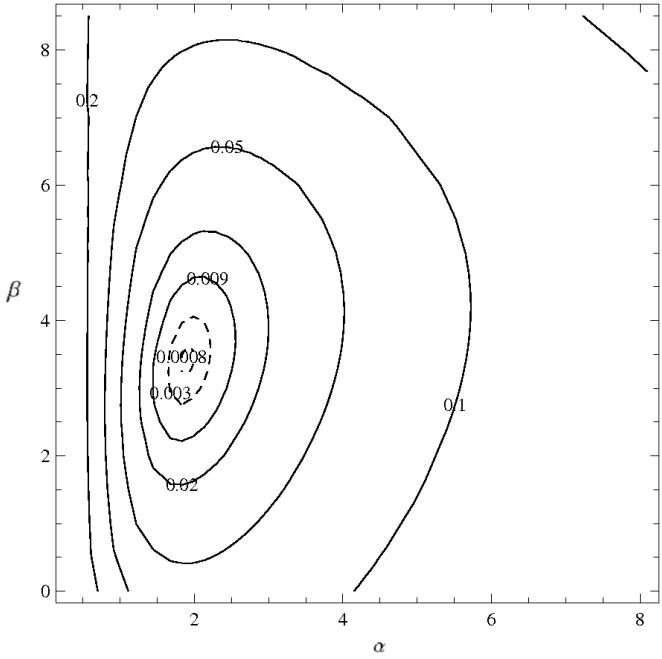}
      \caption{Contours of the function $\Xi$ for the
      case of the self-consistent decomposition presented in this paper.}
         \label{nparametric}
   \end{figure}
 
The results obtained from the standard parametric decomposition
are shown in Fig.~\ref{parametric}, which clearly exhibits the
disk-halo degeneracy pattern. In particular, consider the contour
marked by $0.02$. A sizable diagonal strip in parameter space
consists of points that are basically equivalent from the point of
view of the quality of the fit, even though they correspond to
models that are physically very different, ranging from fairly
light disks up to maximum-disk solutions (with $\beta$ changing by
a factor higher than ten, from $\approx 0.5$ to $\approx 6.5$).
 
In contrast, Fig.~\ref{nparametric} shows the contours of 
function (\ref{deviat}) for the case in which the model rotation
curve is calculated using the self-consistent method presented in
this paper. The plot ranges and the values of the solid
line contours plotted in Fig.~\ref{parametric} and
Fig.~\ref{nparametric} are the same. The self-consistent models
give a better disk-halo decomposition for three distinct reasons:
firstly, the value $\Xi_{bf}= \Xi(\alpha_{bf}, \beta_{bf},1.02)\approx0.00056$ of
the function $\Xi$ corresponding to the best-fit model $(\alpha_{bf}, \beta_{bf})$ is more
than ten times smaller than the value attained by $\Xi$ based on
the simple parametric decomposition ($\Xi_{bf}^{nsc}\approx0.006$); secondly, the area of the
parameter space inside the contour $2\Xi_{bf}\approx0.00112$ (or any other sensible
multiple of $\Xi_{bf}$) is quantitatively smaller, that is the
best-fit model is identified more \textit{sharply}; finally, the
contours of the function $\Xi$ in the self-consistent case do not
show any diagonal trend, thus confirming that the disk-halo
degeneracy is removed.

In conclusion, the best-fit model identified by the
self-consistent analysis is defined by the following values of the
relevant dimensionless parameters
 
\begin{equation}
(\alpha_{bf}, \beta_{bf})\approx(1.9, 3.4)~. \label{bf}
\end{equation}
 
\begin{figure}
   \centering
\includegraphics[width=8.7cm, angle=0]{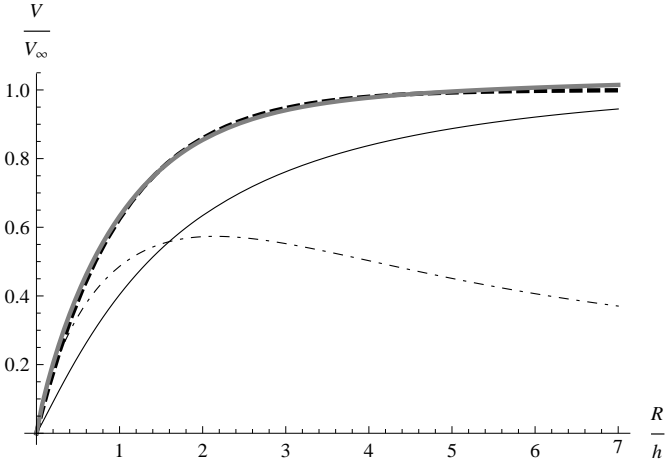}
      \caption{Disk-halo decomposition associated with
      the best self-consistent fit. The thick dashed line
      is the ``observed" rotation curve, the thick solid line
      is the best-fit rotation curve, the thin lines represent
      the disk (dash-dotted) and the halo (solid) contributions.}
         \label{bfrc}
   \end{figure}

\noindent Therefore, the best-fit model is significantly
\textit{sub-maximal} (see Eq.~(\ref{maxdisk})). The associated
rotation curve is shown in Fig.~\ref{bfrc}.
 
In order to better demonstrate the beneficial role of
self-consistency in breaking the disk-halo degeneracy, we now
consider the self-consistent models that correspond to parameters
with good (degenerate) parametric fits to the ``observed" rotation
curve, but are ruled out by the self-consistent analysis. In
particular, as representative cases we choose the high-mass (i.e., high-$\beta$) and
low-mass (i.e., low-$\beta$) ends of the $2\Xi^{nsc}_{bf}$ contour in Fig.~\ref{parametric}:
 
\begin{equation}
\begin{array}{ccl}
(\alpha_{hm}, \beta_{hm})&\approx&(0.9, 5.4)\\ 
(\alpha_{lm}, \beta_{lm})&\approx&(6, 1.75)
\end{array}~.
\end{equation}

\begin{figure}
   \centering
\includegraphics[width=8.7cm, angle=0]{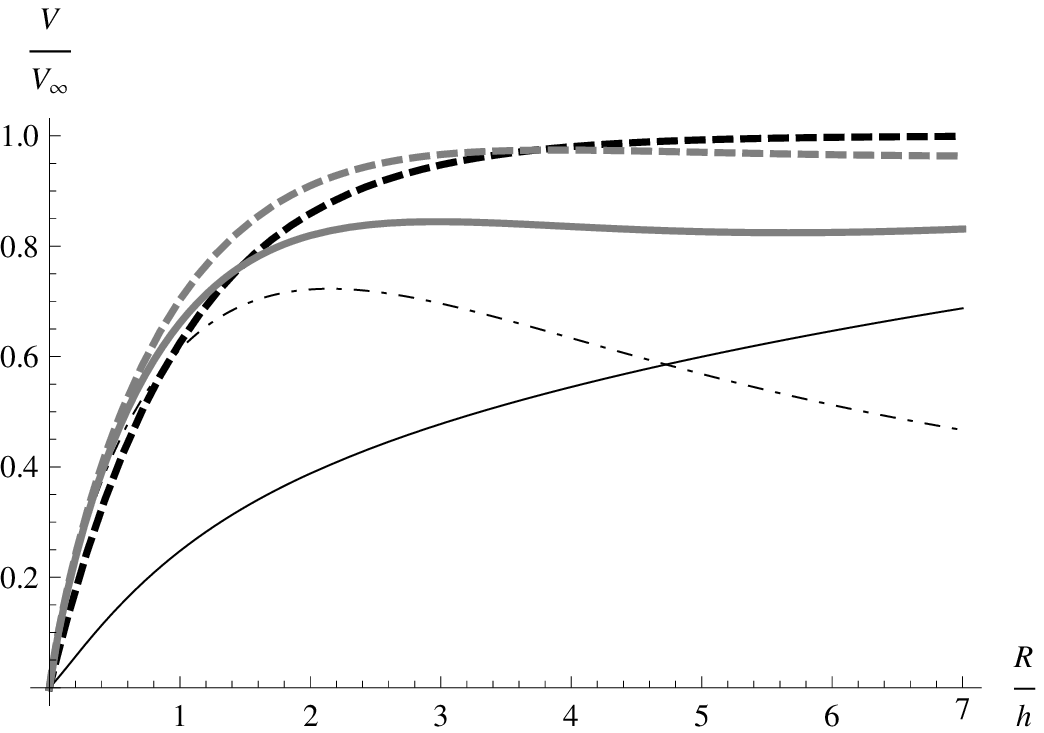}
      \caption{Disk-halo decomposition by the
      $(\alpha_{hm}, \beta_{hm})$ self-consistent model.
      The format and the notation are the same as in Fig.~\ref{bfrc}. The rotation curve of the corresponding $(\alpha_{hm}, \beta_{hm})$ parametric model is displayed as a thick dashed gray line.}
         \label{hmrc}
   \end{figure}
\begin{figure}
   \centering
\includegraphics[width=8.7cm, angle=0]{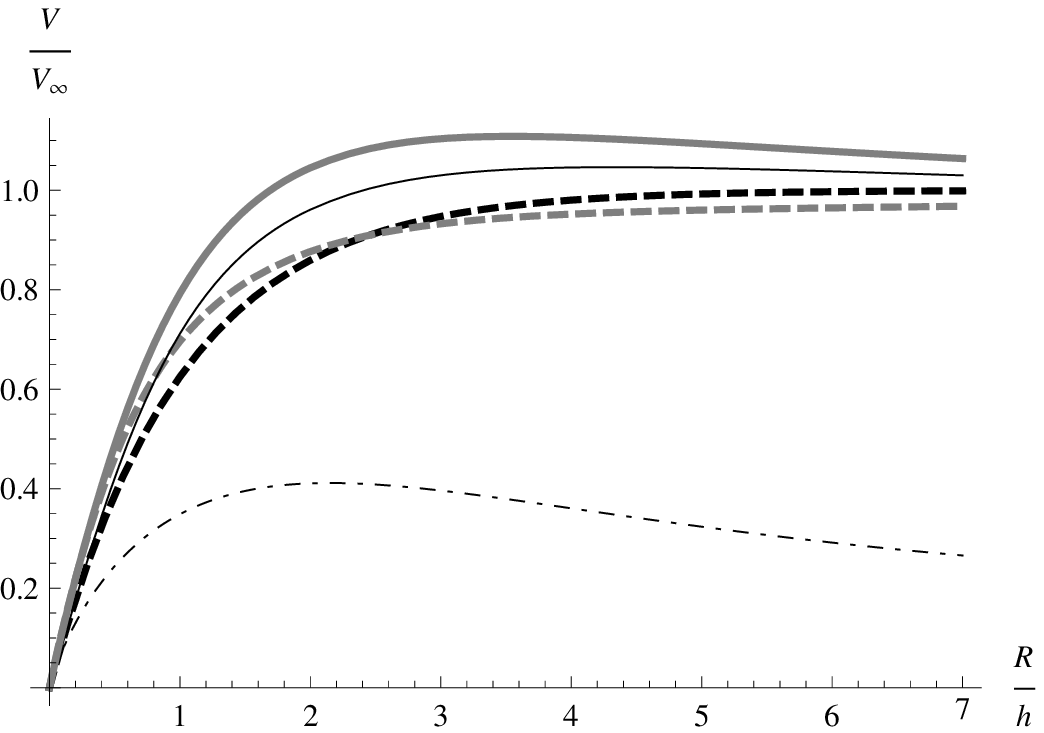}
      \caption{Disk-halo decomposition by the
      $(\alpha_{lm}, \beta_{lm})$ self-consistent model.
      The format and the notation are the same as in Fig.~\ref{bfrc}. The rotation curve of the corresponding $(\alpha_{lm}, \beta_{lm})$ parametric model is displayed as a thick dashed gray line.}
         \label{lmrc}
   \end{figure}

\noindent The self-consistent rotation curves associated with these
two models are shown in Figs.~\ref{hmrc} and \ref{lmrc}. Note that the differences between the calculated total rotation curve and
the ``observed" one are significantly larger than any realistic
error-bar. Furthermore, the rotation curve of the high-mass case confirms the properties of the self-consistent rotation curves described in Sect.~\ref{shaperc}.

\subsection{Application to the case of NGC 3198}\label{3198}

As a specific example, we now consider the classical case of NGC 3198. We refer to the rotation curve reported in Table 2 of van Albada et al. (1985) and to their measure of the exponential length ($h = 2.68$ kpc). The value of the ratio $R_{\Omega}/h$ estimated from the rotation curve is $R_{\Omega}/h = 0.6\pm 0.1$. This is on the low side of the correlation reported in Sect.~\ref {empiric}, that is the rotation curve of NGC 3198 rises more rapidly than average in its central parts. This feature may be partly related to the fact that the photometric profile of NGC 3198 is not exactly exponential in the innermost parts of the disk (see Begeman \cite{begeman89}; see also the discussion by Dutton et al. 2005). For simplicity, in the following we will consider the galaxy as bulgeless and we will ignore the contribution of the gaseous disk. We recall that the simple analysis provided in Sect.~\ref{degremoval}, applied to the inner parts of the rotation curve, suggests that the maximum-disk decomposition for NGC 3198 has
$\beta_{max}^{NGC 3198}\approx 10$.

In this case the rotation curve is provided with its observational errors. To evaluate quantitatively the goodness of a given disk-halo decomposition, we then adopt the classical likelihood function:

\begin{equation}
\Lambda({V_{\infty}}, \alpha, \beta) = \exp\left\{-{1\over 2} \sum_{i=1}^{i=28}{{\left[V_{mod}(R_i)-V(R_i)\right]^2}\over{\mathrm {Var}\left[V(R_i)\right]}}\right\}\ ,
\label{lik}
\end{equation}
\noindent where $\mathrm{Var}\left[V(R_i)\right]$ denotes the square of the error associated with the $i$-th rotation curve data-point. In order to bring out the explicit dependence of the likelihood $\Lambda $ on the assumed value of $V_{\infty}$, we note that the previous equation can be written as:
\begin{equation}\Lambda({V_{\infty}}, \alpha, \beta) = \exp \left\{-{1\over 2} \sum_{i=1}^{i=28}{{\left[V_{\infty}\cdot v_{mod} (R_i,\alpha,\beta)-V(R_i)\right]^2}\over{\mathrm{Var}\left[V(R_i) \right]}}\right\}~,
\end{equation}

\noindent in which the function $v_{mod}(R,\alpha,\beta)$ is normalized to unity at large radii.

We note that the rotation curve data from four to eleven exponential lengths suggest

\begin{equation}
\hat V_{\infty}=(149.8\pm0.5)\ \mathrm{km/s}~.
\label{vinf}
\end{equation}

\noindent However, in order to avoid a possible bias that might be introduced by ``freezing" this quantity, we treat the asymptotic velocity $V_{\infty}$ as a free parameter, to be determined by the fitting procedure together with $\alpha$ and $\beta$.

\begin{figure}
\centering
\includegraphics[width=8.7cm, angle=0]{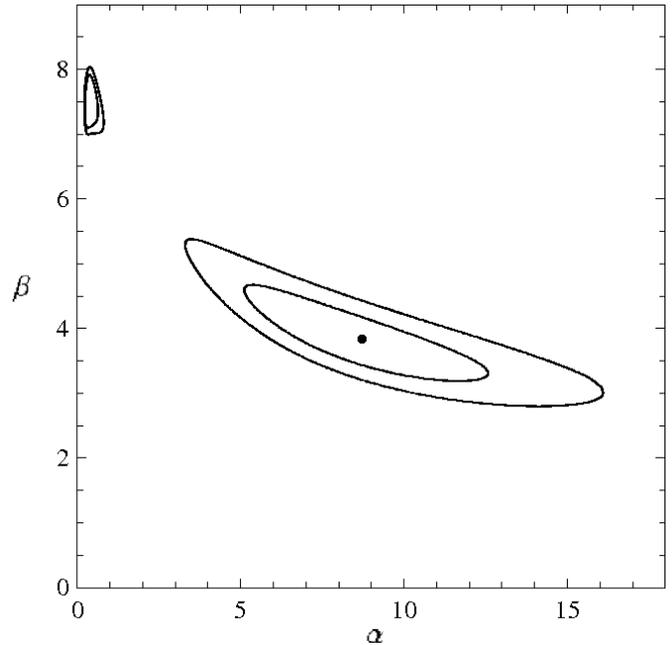}
\caption{Contours of $68\%$ and $95\%$ confidence regions in the $ (\alpha, \beta)$ plane according to the likelihood defined in Eq.~ (\ref{likred}) for the parametric decomposition. The full dot marks the best-fit model.}
\label{p3198}
\end{figure}

Figure~\ref{p3198} shows the results of the parametric disk-halo decomposition in the $(\alpha, \beta)$ plane, that is, after integrating out the dependence on $V_{\infty}$ in the likelihood Eq.~(\ref{lik}):

\begin{equation}\hat\Lambda(\alpha, \beta) =\int_{0}^{\infty} dV_ {\infty}\ \Lambda({V_{\infty}}, \alpha, \beta)~.
\label{likred}
\end{equation}

\begin{figure}
\centering
\includegraphics[width=8.7cm, angle=0]{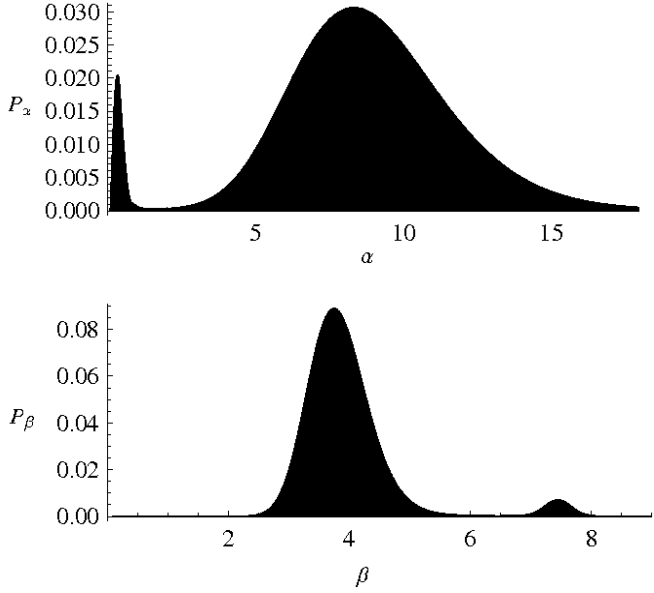}
\caption{Projections of the likelihood (\ref{lik}) for the single parameters $(\alpha,\beta)$ for the parametric decomposition.}
\label{p3198p}
\end{figure}

\begin{figure}
\centering
\includegraphics[width=8.7cm, angle=0]{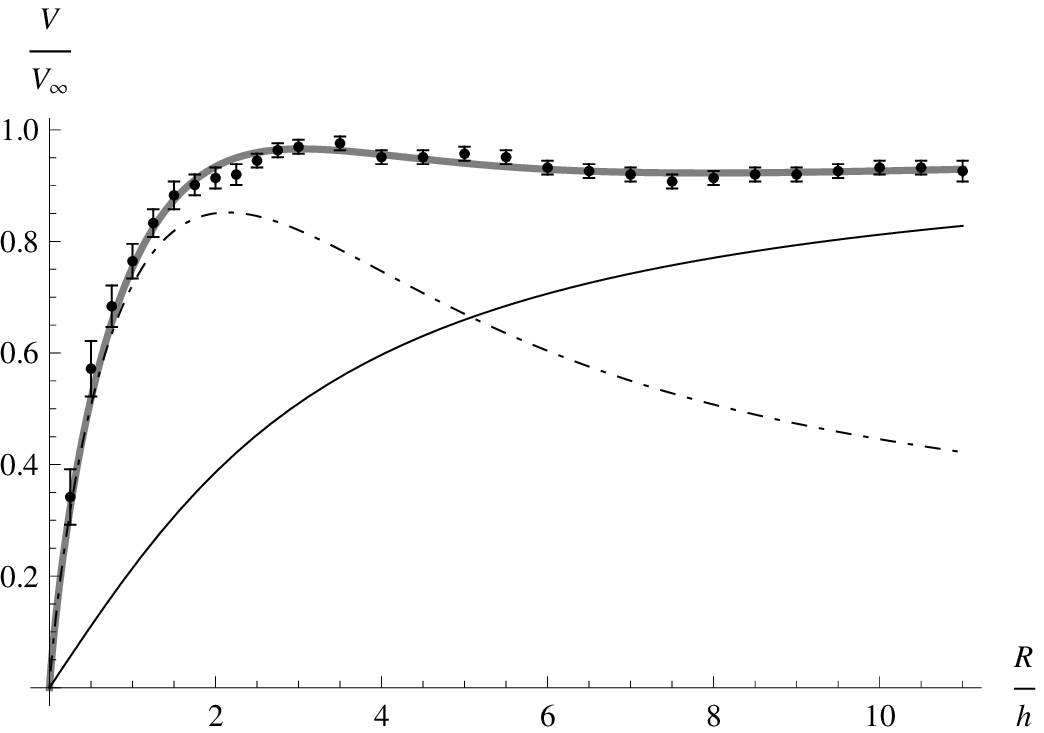}
\caption{Parametric disk-halo decomposition of NGC 3198 corresponding to the maximum of the likelihood in Fig.~\ref{p3198}, with higher disk importance. The coding is the same as in Fig.~\ref{bfrc}.}
\label{max3198}
\end{figure}

\begin{figure}
\centering
\includegraphics[width=8.7cm, angle=0]{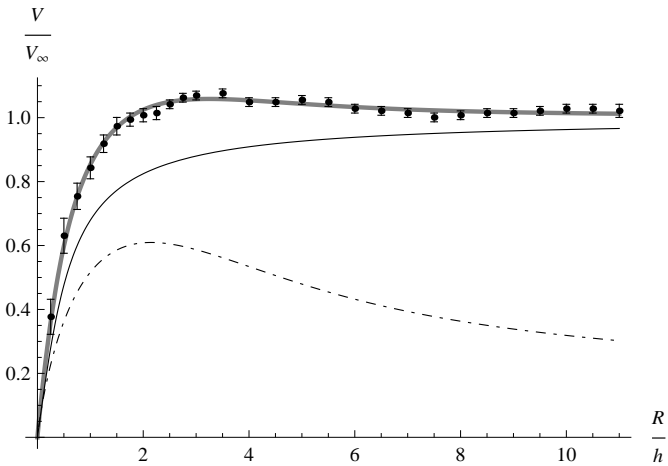}
\caption{Parametric disk-halo decomposition of NGC 3198 corresponding to the maximum of the likelihood in Fig.~\ref{p3198}, with lower disk importance. The coding is the same as in Fig.~\ref{bfrc}.}
\label{min3198}
\end{figure}

\noindent The contours identify regions of $68\%$ and $95\%$ confidence level; the best-fit model is marked by a full dot. Interestingly, two maxima exist, located along the disk-halo degeneracy strip described in the previous subsection, but with significantly different physical properties. In fact, for varying values of the asymptotic velocity $V_{\infty}$, the location of the best-fit model in the $(\alpha, \beta)$ plane (at fixed $V_{\infty}$) changes. As a result, the integrated likelihood Eq.~(\ref{likred}) actually increases the amount of degeneracy associated with the parametric disk-halo decomposition. This is particularly evident from Fig.~\ref {p3198p}, which shows the projections along the two separate parameters $\alpha$ and $\beta$ of the the likelihood itself, exhibiting a clear bimodality. The disk-halo decomposition is unable to select between a solution very close to the maximum-disk and a decomposition with a much lighter disk. The corresponding decompositions are illustrated in Figs.~\ref{max3198} and \ref {min3198}, which show the two best-fit parametric decompositions, with projections on the $(\alpha, \beta)$ plane contained respectively in the two separate areas of the confidence regions in Fig.~\ref{p3198}. This curious result may be related to the specific structure of the rotation curve of NGC 3198 (but see also Appendix B). In any case, the general statement holds, that the parametric decomposition is unable to lead to a unique determination of the weight of the stellar disk.

For completeness, we record the coordinates of the best-fit model in the $(V_{\infty}, \alpha, \beta)$ space, together with the related reduced $\chi^2$ value:

 \begin{equation}
(145.9, 8.7, 3.85, 0.6)~.
 \end{equation}

\noindent Here we have defined

\begin{equation}
\chi^2_{red}={1\over {28-\textbf{3}}}\sum_{i=1}^{i=28}{{\left[V_{mod} (R_i)-V(R_i)\right]^2}\over{\mathrm{Var}\left[V(R_i)\right]}}~.
\end{equation}


\begin{figure}
\centering
\includegraphics[width=8.7cm, angle=0]{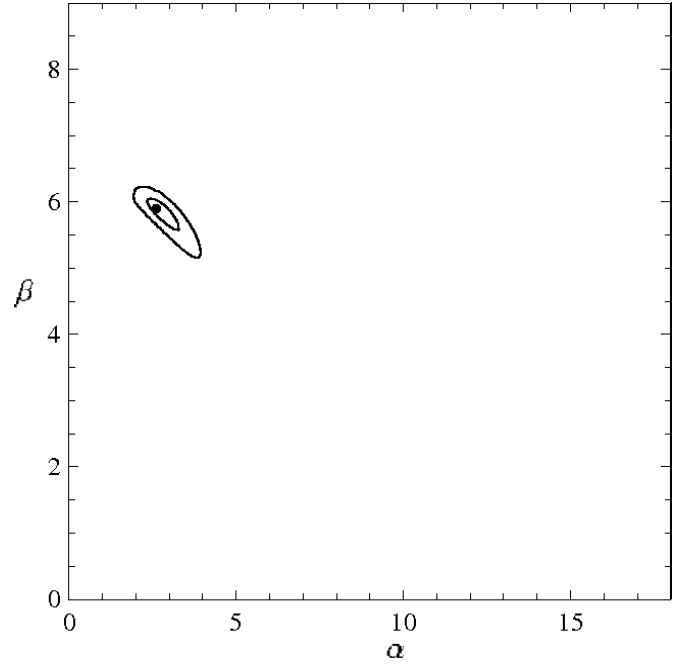}
\caption{Contours of $68\%$ and $95\%$ confidence regions in the $ (\alpha, \beta)$ plane according to the likelihood defined in Eq.~ (\ref{lik}) for the self-consistent decomposition. The full dot marks the best-fit model.}
\label{nnp3198}
\end{figure}

\begin{figure}
\centering
\includegraphics[width=8.7cm, angle=0]{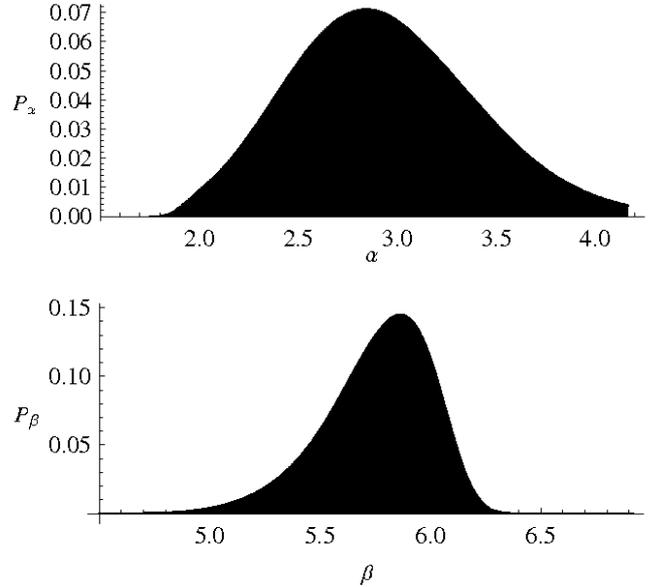}
\caption{Projections of the likelihood (\ref{lik}) for the single parameters $(\alpha,\beta)$ for the self-consistent decomposition.}
\label{nnp3198p}
\end{figure}

The situation is completely different if the disk-halo decomposition is performed by means of the self-consistent models presented in this paper. Figure~\ref{nnp3198} shows the contours of the $68\%$ and $95\% $ confidence levels and the best-fit model, calculated using the same likelihood as for the parametric case; note the significant change in their size compared to Fig.~\ref{p3198}. At fixed values of the asymptotic velocity $V_{\infty}$, the general behavior of the best-fit models is similar to the one noted for the parametric analysis: the higher the adopted value of $V_{\infty}$, the higher the value of $\beta$ found in the best-fit decomposition. However, here the differences cancel out in the integration Eq.~(\ref{likred}), because of the different sharpness of the best-fits obtained at fixed values of $V_{\infty}$. As a result, the self-consistent decomposition is determined uniquely. In Fig.~\ref {nnp3198p} the likelihood projections are indeed smooth and do not exhibit a bimodal shape. As for the parametric decomposition, we list the coordinates of the best-fit model in the $(V_{\infty}, \alpha, \beta)$ space, together with the related reduced $\chi^2$ value:

\begin{equation}
(160.1, 2.6, 5.9, 0.7)~.
\end{equation}
\noindent The best-fit value of the asymptotic velocity $V_{\infty}$ is  higher than the value suggested by the rotation curve in Eq.~(\ref{vinf}), consistent with the properties of the rotation curves displayed in Fig.~\ref{panels} and with the \textit{suppression effect} described in Sect.~\ref{shaperc}.


\begin{figure}
\centering
\includegraphics[width=8.7cm, angle=0]{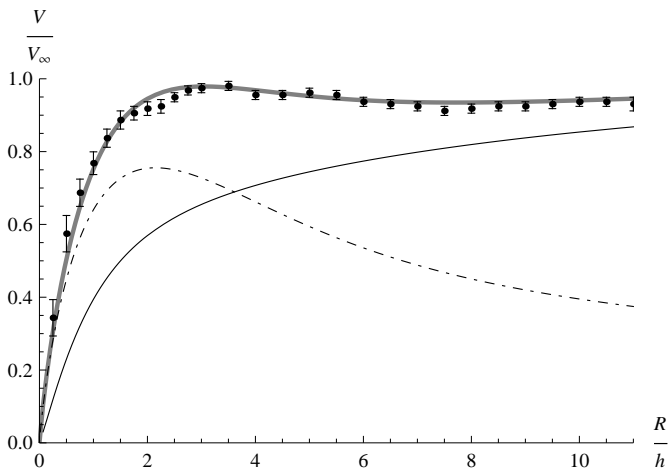}
\caption{Self-consistent disk-halo decomposition of NGC 3198. The coding is the same as in Fig.~\ref{bfrc}.}
\label{fit3198}
\end{figure}

We note that the reduced $\chi^2$ value of the best self-consistent fits is somewhat higher than for the corresponding parametric fits. However, this difference is small and the reported values for the reduced $\chi^2$ account for globally good fits. This is evident from Fig.~\ref{fit3198}, which shows the best-fit self-consistent disk-halo decomposition together with the observed rotation curve. However, the model has a certain difficulty in fitting the second and third data-points, which remain higher than the best-fit self-consistent rotation curve. Clearly, this could be ascribed to the exponential modeling of the density profile of the visible matter, which, as noticed before, does not take into account the small bulge and the gas present in NGC 3198. Since these two data-points are those that define the value of the ratio $R_{\Omega}/h$ for NGC 3198, Fig.~\ref{figromega} confirms that the best-fit self-consistent model has a slightly higher value for this ratio. Note also that the properties of the best-fit self-consistent decomposition are quite different from those found in the parametric analysis, with the best-fit model characterized by a disk of intermediate weight.  As is evident from Fig.~\ref{fit3198}, the disk contribution to the rotation curve is important in the inner parts of the system. However, with its $\beta\approx 6$, the best-fit self-consistent model remains below the classical maximum-disk solutions that van Albada et al. (1985), for example, place at $\beta\gtrapprox 9$.

\section{Conclusions}
\label{Sect:Discussion}

This paper has addressed the construction of self-consistent
models for a dark matter distribution assumed to be isothermal, in
the presence of a zero-thickness disk. The models have then been
applied to interpret the properties of observed rotation curves,
with special attention to the issues of conspiracy and degeneracy
that are known to affect their disk-halo decomposition.

We argue that what has been learned from the study
of the strictly isothermal case should be applicable to the
case of dark matter distributions characterized by a quasi-isothermal
inner region (as is the case for the results of collisionless
violent relaxation).

It would be interesting to test the results on the dark matter distributions predicted by current cosmological models, which definitely favor a density profile asymptotic to $r^{-3}$. This test could be performed with an approach similar to the one adopted in this paper, provided that we had a physically justified distribution function to start from. In particular, it would be interesting to test to what extent the cosmologically predicted distributions could be reconciled with the existence of cases, such as that of NGC 3198, in which the rotation curve remains very flat in a wide radial range.

Note that this paper has considered the impact of a \textit{given} visible matter distribution on the distribution of dark matter. A study of the dynamical role of the dark halo in shaping the distribution of the visible matter would require considering a formation scenario that we do not wish to address here.

The main results obtained in this paper are the following:

\begin{itemize}

\item{As a preliminary empirical study, we have shown that \textit{for both high surface brightness
and low surface brightness disk galaxies} the scale length $h$ of
the light profile of pure disks correlates with the scale length
$R_{\Omega}$ (defined in Eq.~(\ref{romega})) that characterizes
the central growth of the rotation curve. This suggests that the
distribution of dark and visible matter in pure disk galaxies is
homologous, i.e. characterized by similar global relations in different galaxies, even if significantly different in size and type (HSB vs. LSB galaxies). Based on the results obtained in the rest of the
paper, this point is interpreted as an indication of a
gravitational interplay between luminous and dark components in
the innermost parts of the system. We argue that this homology is
more relevant and surprising as a conspiracy than the actual
flatness of the observed rotation curves, because, especially in
the context of parametric models, rotation curves turn out to be
almost flat for very different values of the disk-to-halo mass
ratio.}

\item{We have constructed self-consistent models of isothermal halos
embedding zero-thickness stellar disks, thus obtaining fully
self-consistent axisymmetric isothermal dark matter halos. These
make basically a two-parameter family of models; the two
parameters measure the weight of the disk and the central density
of the halo (see Eqs.~(\ref{alpha}, \ref{beta})). A third parameter,
 the asymptotic value of the rotation curve at large radii, 
is in principle free but can often be treated as a fixed parameter, 
given the fact that it is generally well constrained by 
the observations; note that a discussion with the third 
parameter kept free is given in our study of NGC 3198 (see fifth item in 
this subsection). The adopted solution procedure can easily 
incorporate the presence of a zero-thickness
disk of gas (with desired surface density). In addition, a similar
procedure could be followed to construct models based on different
choices of the dark matter distribution function. In turn, we can
also argue that the qualitative conclusions obtained in this paper
by calculating the properties of strictly isothermal halos should
also apply to more general systems where the halo distribution, in
the region of interaction with the disk, is quasi-isothermal.}

\item{We have shown and quantified that, for astrophysically realistic values of the
two free parameters of the models, the dark matter halo is
significantly flattened, especially in the regions where it
coexists with the bright optical disk. Inside one exponential scale
 length, the dark matter distribution is that of a thick disk;
 nevertheless, its contribution to the rotation curve is approximately 
linear in the same region, denoting the presence of a natural core.}

\item{The rotation curves of the self-consistent models
have been shown to be generally smooth and flat in their
intermediate regions. In particular, self-consistency has been
proved to provide a physical ``mechanism" able to flatten rotation
curves in the region of the visible disk. This \textit{flattening
effect} is stronger in the region of parameter space that appears
to represent realistic systems. This result goes against the naive
expectation that self-consistency should tend to enhance features
present in the potential well and then in the related rotation
curve.}

\item{We have shown that the application of self-consistent models
to decompose an observed rotation curve breaks the disk-halo
degeneracy. This result has been achieved under some simplifying assumptions. First, for simplicity we assumed that the density profile of the disk is a perfect exponential (although our self-consistent models can be constructed for any zero-thickness axysimmetric density profile for the stellar disk). Second the dark matter distribution function has been assumed to be a perfect isothermal; this point can be released in favor of other physically justified choices. The disk-halo degeneracy breaking
has been obtained first by studying an idealized rotation curve representative of a typical rotation curve of a pure disk spiral galaxy and then by a direct
application to the classical case of NGC 3198. We have also
provided arguments that support the idea that, to some extent,
such degeneracy problem has been artificially created by the use
of parametric models (such as the model associated with
Eq.~(\ref{vdm})). In fact, the rotation curves generated by these
parametric models have been shown to be exceedingly uniform and
featureless.}

\item{We have thus derived the best-fit self-consistent disk-halo
decompositions and proved that they favor mass decompositions
characterized by stellar disks of intermediate weight, which are
still important in the central regions of the system, but are
significantly lighter than maximum-disk solutions. On the other hand, the homology property noted in the first item of these Conclusions would still make such sub-maximal solutions consistent with the small scatter of the Tully-Fisher (1977) relation. Of course, a full demonstration of this conclusion would require that we moved beyond the case of the fiducial rotation curve (see Sect.~5.1),  by studying, in the way described in Sect.~5.2 for NGC 3198, a complete sample of spiral galaxies. But this would obviously go well beyond the objectives of the present paper.}

\item{The resulting
rotation curve has been interpreted as resulting from a
\textit{suppression effect}, characteristic of self-consistency.
Self-consistency would affect a proposed maximum-disk
decomposition by significantly reducing the dark halo contribution 
to the total rotation curve and thus leading to worse fits, especially 
at intermediate radii. }

\item{The resulting $M/L$ values for the stellar disk should be studied for each individual galaxy, in view of their implications and consistency with respect to what is known about the relevant stellar populations. In this context, an important test would be the study of LSB galaxies. However, this very important issue goes well beyond the goals of the present paper.}

\end{itemize}

In a follow-up paper we will generalize the results of this paper
to the case in which the disk is characterized by finite
thickness, with results that may be of interest to projects such
as the Disk Mass Project (Verheijen et al. 2004 and Westfall et al. 2008).

\begin{acknowledgements}
We wish to thank Luca Ciotti, Wyn Evans, Renzo Sancisi, Tjeerd van Albada and an anonymous Referee for making a number of interesting comments and suggestions.
\end{acknowledgements}

\appendix

\section{Construction of the models by an iterative procedure} \label{numsol}
Equation (\ref{poishalo}), with the boundary conditions
(\ref{boundary}), is solved by means of a modified version of the
iterative procedure presented earlier by Prendergast \& Tomer (1970) in view of a completely different problem (the
study of nonspherical models for rotating elliptical galaxies).
It is based on the classical expansion in Legendre polynomials, a
natural approach in the case of axisymmetric systems. However, the
original method is viable only for gravitational potentials that
are regular at large radii. In contrast, in our model, the
potential of the isothermal halo diverges at large radii. This
difficulty has been overcome by separating the potential of the
dark matter halo into two different parts: a regular part, which
converges to finite values at large radii, and a divergent one,
which follows the asymptotic prescription of Eq.~(\ref{boundary}).
Below we provide the details of the technique developed in the
present paper.
Here it is convenient to work with standard spherical coordinates
$(r, \phi, \theta)$ and their dimensionless analogue $(\eta, \phi,
\theta)$. We start by recalling briefly the method devised by
Prendergast \& Tomer (1970). The solution of the two-dimensional Poisson
equation
\begin{equation}
\nabla^2\psi(\eta,\theta)=\hat\rho\left[\eta, \theta; \psi(\eta, \theta)\right]
\label{pois2d}
\end{equation}
\noindent is obtained by iteratively solving the equations
\begin{equation}
\nabla^2\psi^{(n+1)}(\eta,\theta)=\hat\rho\left[\eta, \theta; \psi^ {(n)}(\eta, \theta)\right]
\label{poisiter}
\end{equation}
\noindent to which the same boundary conditions as in
Eq.~(\ref{pois2d}) are imposed. The potential at the iterative
step $(n+1)$, $\psi^{(n+1)}$, is obtained from the potential at
the previous step, $\psi^{(n)}$, by solving Eq.~(\ref{poisiter})
exactly, using the standard multipole expansion in Legendre
polynomials, here denoted by $P_k$:
\begin{equation}
\hat\rho^{(n)}(\eta,\theta)=\sum_{k=0}^{\infty}\
\hat\rho^{(n)}_{k}(\eta)\ P_{k}(\cos\theta)~, \label{densexp}
\end{equation}
and
\begin{equation}
\psi^{(n+1)}(\eta,\theta)=\sum_{k=0}^{\infty}\
\psi^{(n)}_{k}(\eta)\ P_{k}(\cos\theta)~. \label{potexp}
\end{equation}
\noindent For completeness, we display the general solution of the
$n$-th iterative step:
\begin{equation}
\begin{array}{ll}
\psi^{(n+1)}(\eta, \theta)=& \Psi
+\left[\int_{0}^{\eta}\eta'\hat\rho^{(n)}_{0}(\eta')d\eta'-{1\over \eta}\int_{0}^{\eta}\eta'^2\hat\rho^{(n)}_{0}(\eta')d\eta'
\right]-\\
&-\sum_{k=2}^{\infty}{{P_k(\cos\theta)}\over{2k+1}}\left[\eta^k \int^{\infty}_{\eta}\eta'^{1-k}\hat\rho^{(n)}_{k}(\eta')d\eta'+ \right.\\
&\left.+{1\over{\eta^{k+1}}}\int_{0}^{\eta}\eta'^{k+2}\hat\rho^ {(n)}_{k}(\eta')d\eta' \right]
\end{array}
\end{equation}
\noindent where we have used the notation $\psi(0, \theta)=\Psi$.
The iteration can be stopped when the desired accuracy
prescription is met, as for example when in the relevant domain
$\mathcal{D}$ we find that
\begin{equation}
\max_{(\xi,\zeta)\in\mathcal{D}}\left|{{\psi^{(n+1)}
-\psi^{(n)}}\over{\psi^{(n+1)}+\psi^{(n)}}}\right|<\epsilon~.
\label{erriter}
\end{equation}
The dark matter gravitational potential is then separated into two
parts:
\begin{equation}
\psi_{DM}\equiv \psi_{asy}+\psi~. \label{separation}
\end{equation}
\noindent The potential defined as $\psi_{asy}$ obeys the Poisson
equation
\begin{equation}
\nabla^2\psi_{asy}=\hat\rho_{asy}
\label{asysep}
\end{equation}
\noindent and has the same asymptotic behavior as in
Eq.~(\ref{boundary}). In this way the potential $\psi$ converges
to zero at large radii and the related Poisson equation
\begin{equation}
\left({1\over\xi}{\partial\over{\partial\xi}}\xi{\partial\over {\partial\xi}}
+{\partial^2\over{\partial\zeta^2}}\right)\psi=-\alpha
\exp{\left[\psi_{asy}+\psi_{D}+\psi\right]}- \hat\rho_{asy}
\label{final}
\end{equation}
\noindent can be directly solved by the iterative multipole
expansion outlined previously.

Obviously, there is an infinite number of pairs $(\hat\rho_{asy}, \psi_ {asy})$ that meet the conditions of Eqs.~(\ref{asysep}) and (\ref {boundary}). Therefore, the construction method proposed below just reflects one reasonable choice. We decided to take $\hat\rho_{asy}$ and $\psi_{asy}$ with spherical symmetry, because the relevant asymptotic condition for $\psi_{asy}$ is characterized by spherical symmetry and because the spherical Poisson equation admits a simple explicit solution:

\begin{equation}
\psi_{asy}(\eta)=\psi_{asy}(0)+\int_{0}^{\eta}\eta'\hat\rho_ {asy}(\eta')d\eta'-{1\over\eta}\int_{0}^{\eta}\eta'^2\hat\rho_{asy} (\eta')d\eta' \ .
\label{spherical}
\end{equation}

\noindent The precise choice of the form of the density $\hat\rho_ {asy}$ is guided by the goal of simplifying the following numerical procedure to solve Eq.~(\ref{final}). Different choices of the density $\hat\rho_{asy}$ correspond to different shapes of the potential $\phi$ for the same pair of free parameters $(\alpha, \beta) $. The adopted practical recipe to construct the density $\hat\rho_ {asy}$ turns out to be useful and efficient.

We start by defining what we call an ``observed" pseudo-potential, constructed from the observed rotation curve:

\begin{equation}
\Phi_{obs}(r)=\int_{0}^{R}{{V^2(s)}\over{s}}ds\ .
\end{equation}

\noindent At large radii it has the asymptotic expression

\begin{equation}
\Phi_{obs}(r)\sim V^2_{\infty}\ln\left({r\over{r_0^{obs}}}\right)\ ,
\label{apsiobs}
\end{equation}

\noindent where the radius $r_0^{obs}$ can be calculated from the precise form of the rotation curve. The behavior of Eq.~(\ref {apsiobs}) naturally relates the potential $\Phi_{obs}$ (and its dimensionless counterpart $\psi_{obs}$) to a density distribution with isothermal functional shape:

\begin{equation}
{1\over{\eta^2}}{d\over{d\eta}}\eta^2{d\over{d\eta}}\psi_{obs}(\eta) \sim-\alpha_{obs}\exp\left[\psi_{obs}(\eta)\right]\ ;\ \eta\gg1\ .
\label{alphaobs}
\end{equation}

\noindent As shown in Eq.~(\ref{asycon}), the value of the constant $ \alpha_{obs}$ is determined by the value of the radius $r_0^{obs}$ only, and is thus fixed. Having made this points clear, we can define $\hat\rho_{asy}$ to be

\begin{equation}
\hat\rho_{asy}(\eta)\equiv-\alpha_{obs}\exp\left[\psi_{obs}\left(\sqrt {\eta^2+\eta_c^2}\right)\right]+\hat\rho_D^{(0)}(\eta)\ ,
\label{dens1}
\end{equation}

\noindent where we have introduced a regularizing core structure with characteristic size $\eta_c$ and subtracted the monopole term of the stellar density

\begin{equation}
\rho_D^{(0)}(\eta)\equiv{\beta\over{2}}\int^{1}_{-1}\hat{\Sigma}(\eta) \ \delta(\zeta)\ P_0(\cos\theta)\ d\cos\theta={\beta\over{2}}{{\hat {\Sigma}(\eta)}\over\eta}\ .
\label{monodisk}
\end{equation}

With this density profile, from Eq.~(\ref{spherical}) we calculate the gravitational potential $\psi_{asy}$, using the free constant $ \psi_{asy}(0)$ to make it meet exactly the asymptotic behavior required by Eq.~(\ref{boundary}). The determination of the correct value of the constant $\psi_{asy}(0)$ determines also the boundary condition to be used for Eq.~(\ref{final}), which is set by the prescription:

\begin{equation}
\psi_T(0,0)=\psi(0,0)+\psi_{asy}(0)+\psi_{D}(0,0)=0\ .
\end{equation}

\subsection{The truncation of the angular expansion}

A code has been written to solve Eq.~(\ref{final}) through the technique described in the previous section. The code is able to manage automatically the appropriate number of multipole terms to meet a required accuracy (see Eq.~(\ref{erriter})). At each iterative step, the relevant number of multipole orders is calculated from the following prescription:

\begin{equation}
{{\psi_{\bar{k}}^{(n)}(\bar\eta)\ P_{\bar{k}}(\cos\bar\theta)}\over {\sum_{k=0}^{\bar{k}-1}\psi_{{k}}^{(n)}(\bar\eta)\ P_{{k}}(\cos\bar \theta)}}>10^{-1}\epsilon\ .
\label{nummult}
\end{equation}

\noindent In words, at the $n$-th iteration step, the multipole term of order $\bar k$ is retained in the expansion if its (relative) contribution to the gravitational potential $\psi^{(n)}$ (calculated with the previous $\bar k -1$ multipole orders only) is comparable to the accuracy level $\epsilon$ we want to meet. This comparison is made at the coordinates $(\bar\eta, \bar\theta)$, chosen in the region where the deviation from spherical symmetry is strongest: $\bar\theta\approx\pi/2$ and $\bar\eta\approx 1\div 2$.

The models described in the paper are calculated with an accuracy prescription of $\epsilon=10^{-4}$. Obviously, the number of iterations required by such precision depends on the gravitational importance of the stellar disk (which determines the flattening of the halo). For values of the pair $ (\alpha, \beta)$ that represent an astrophysically realistic configuration (see Sect.~\ref{Sect:Properties}), the number of required iterations is $10\div20$, with a number of multipole orders of $12\div20$.

\section{Anomalous rotation curves }\label{wrongcorr}

In this Section we will analyze the case of rotation curves with inner gradients that are significantly different from the one identified by Eq.~(\ref{corr}). These rotation curves represent systems with a ``wrong tuning" between the disk and halo components, in the sense described in Sect.~\ref{Sect:Conspiracy}. We adopt the same simple parametrization for the rotation curve shape as in Eq.~(\ref{paramprof}) and consider two different values for the parameter $\tau$, taken to be significantly far from the value which that reproduces the correlation in Eq.~(\ref{corr}). On the fast rising end, we take $\tau_{f}=0.4$ (i.e. $R_{\Omega}/h\approx0.44$), while on the opposite side of a slowly rising rotation curve we take $\tau_{s}=2.3$ (i.e. $R_{\Omega}/h\approx2.53$). We apply the self-consistent and the parametric disk-halo decomposition to these two anomalous rotation curves and refer to the same function $\Xi$ as in Eq.~(\ref{deviat}) to quantify the quality of the fit. 

\begin{figure}
   \centering
\includegraphics[width=8.7cm, angle=0]{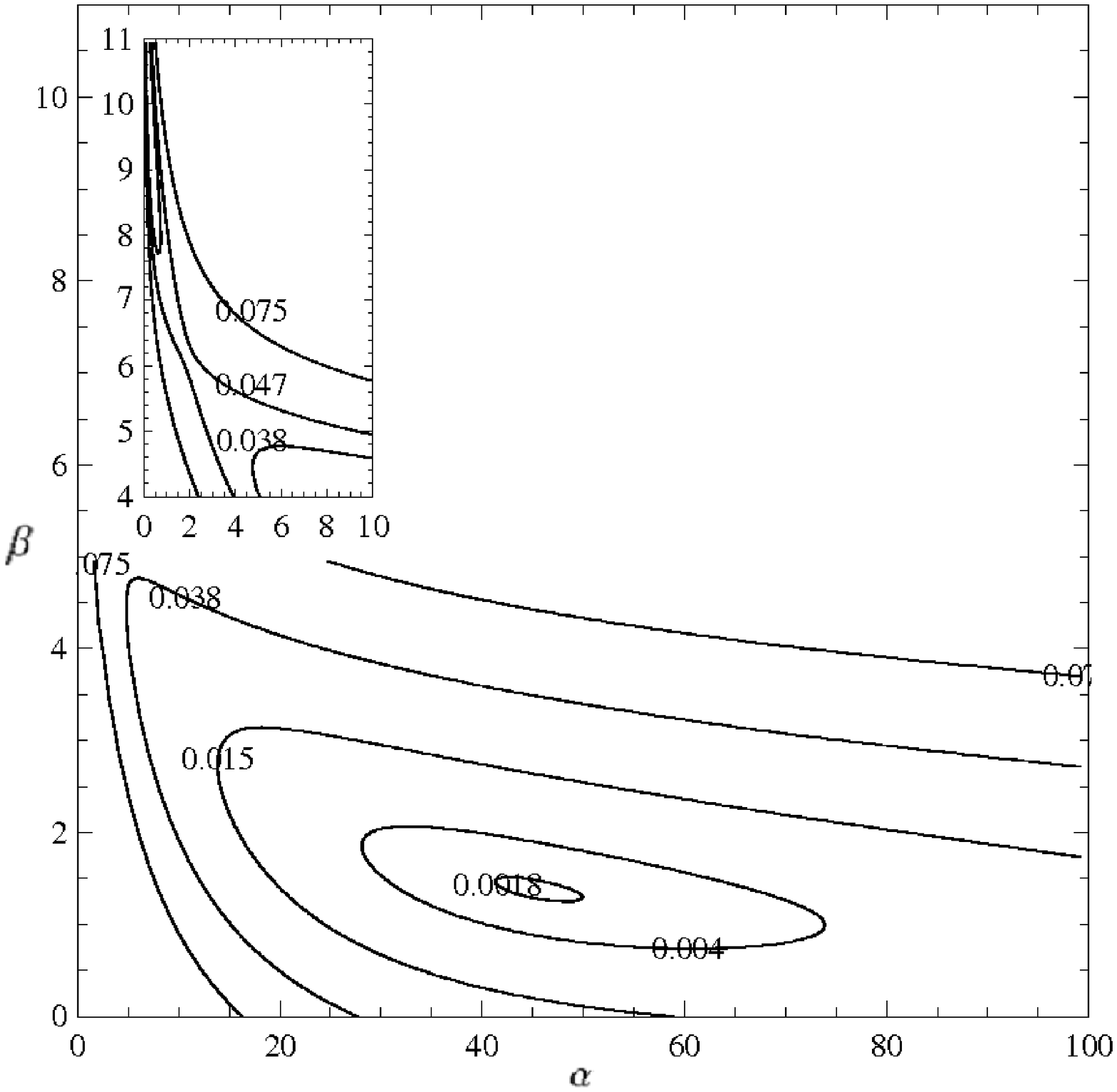}
      \caption{Contours of the function $\Xi$ for the
       parametric decomposition (based on Eqs.~(\ref{vd}, \ref{vdm})) of the rotation curve defined by Eq.~(\ref{paramprof}) with  $\tau_{f}=0.4$.}
         \label{ph}
   \end{figure}
\begin{figure}
   \centering
\includegraphics[width=8.7cm, angle=0]{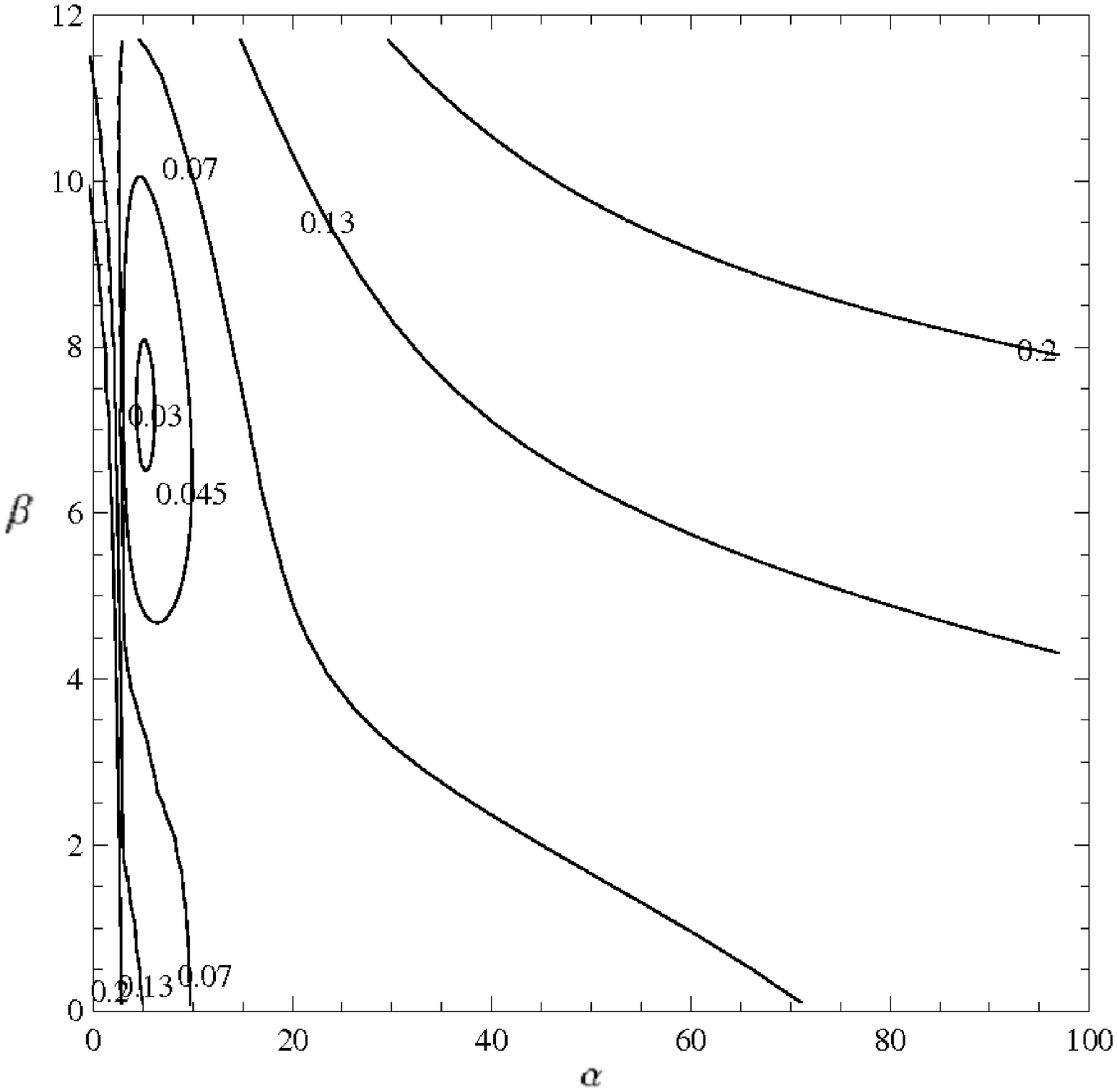}
      \caption{Contours of the function $\Xi$ for the self-consistent decomposition of the rotation curve defined by Eq.~(\ref{paramprof}) with  $\tau_{f}=0.4$.}
         \label{nnph}
   \end{figure}
\begin{figure}
   \centering
\includegraphics[width=8.7cm, angle=0]{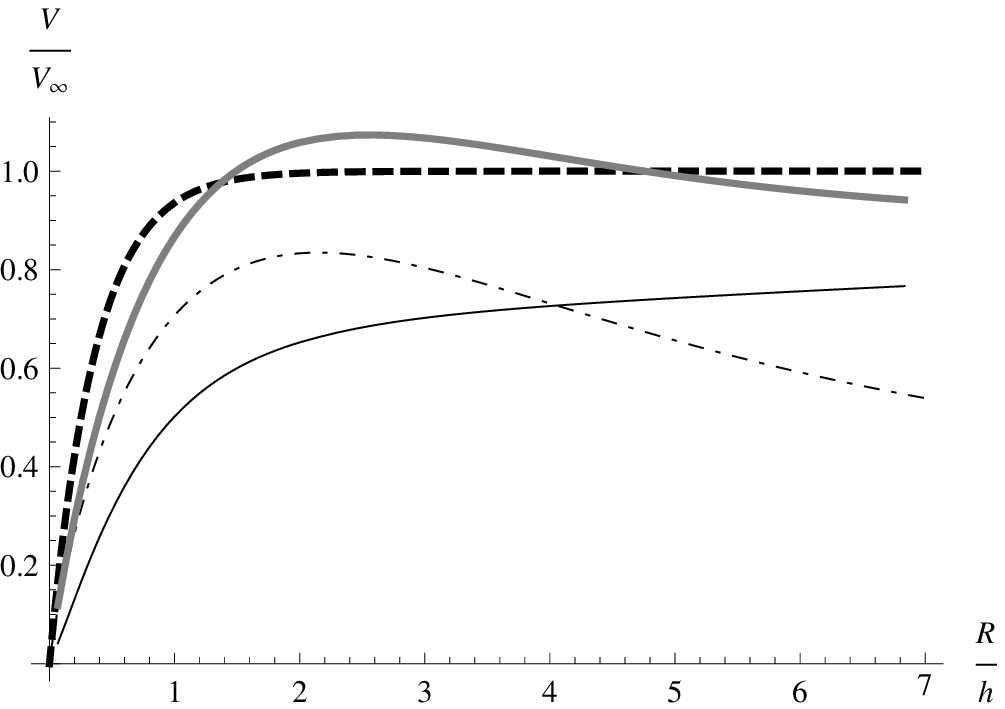}
      \caption{Disk-halo decomposition associated with
      the best self-consistent fit of the case $\tau_{f}=0.4$. The coding is the same as in Fig.~\ref{bfrc}.}
         \label{hrc}
   \end{figure}

The results for the fast rising rotation curve are shown in Fig.~\ref{ph} and Fig.~\ref{nnph}. In the parametric decomposition method we find again the bimodality of the fit, as noted in the case of NGC 3198. The self-consistent method instead identifies a decomposition with an important stellar disk; the fit for the rotation curve is illustrated in Fig.~\ref{hrc}.

\begin{figure}
   \centering
\includegraphics[width=8.7cm, angle=0]{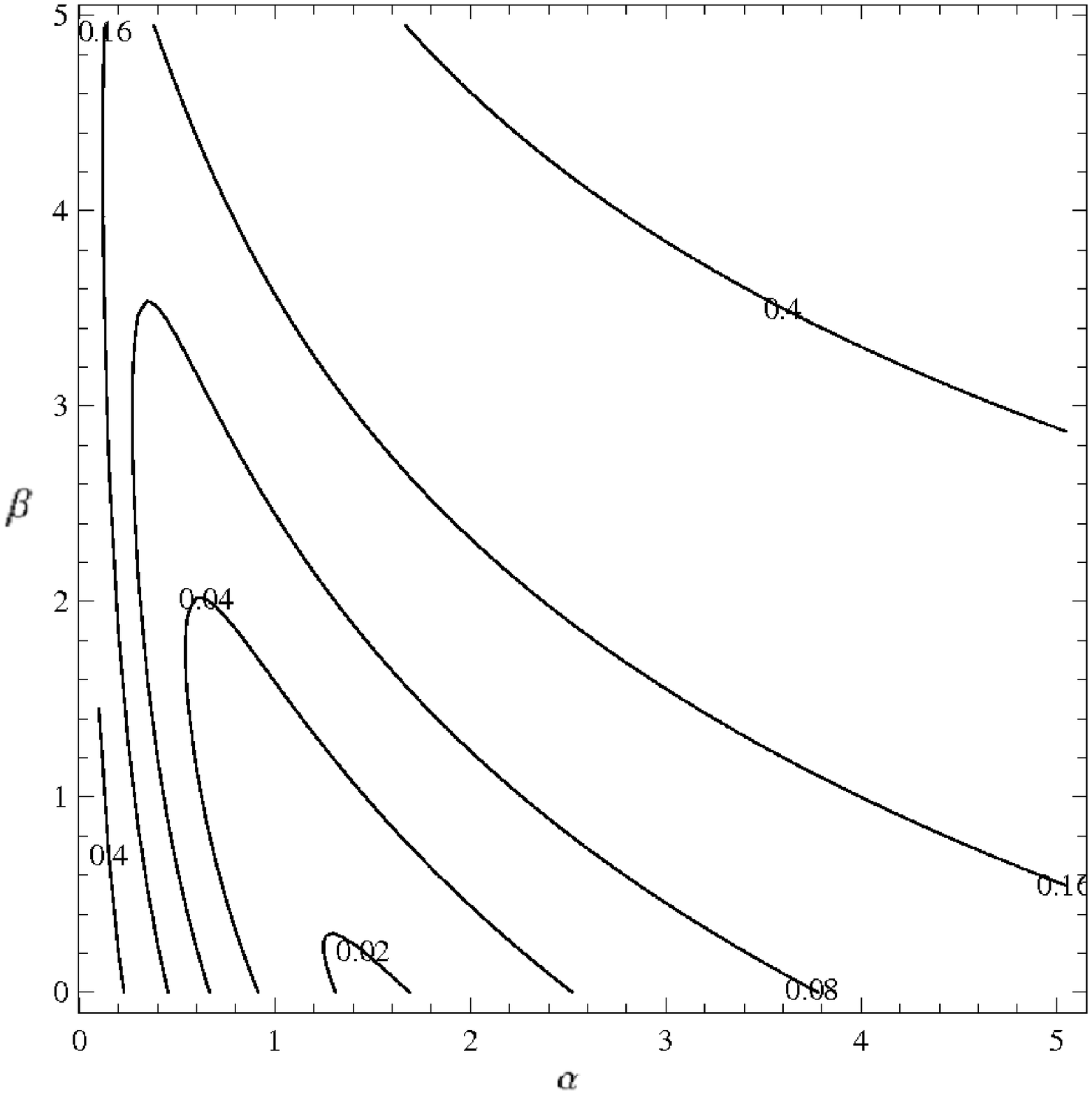}
      \caption{Contours of the function $\Xi$ for the
       parametric decomposition (based on Eqs.~(\ref{vd}, \ref{vdm})) of the rotation curve defined by Eq.~(\ref{paramprof}) with  $\tau_{s}=2.3$.}
         \label{pl}
   \end{figure}
\begin{figure}
   \centering
\includegraphics[width=8.7cm, angle=0]{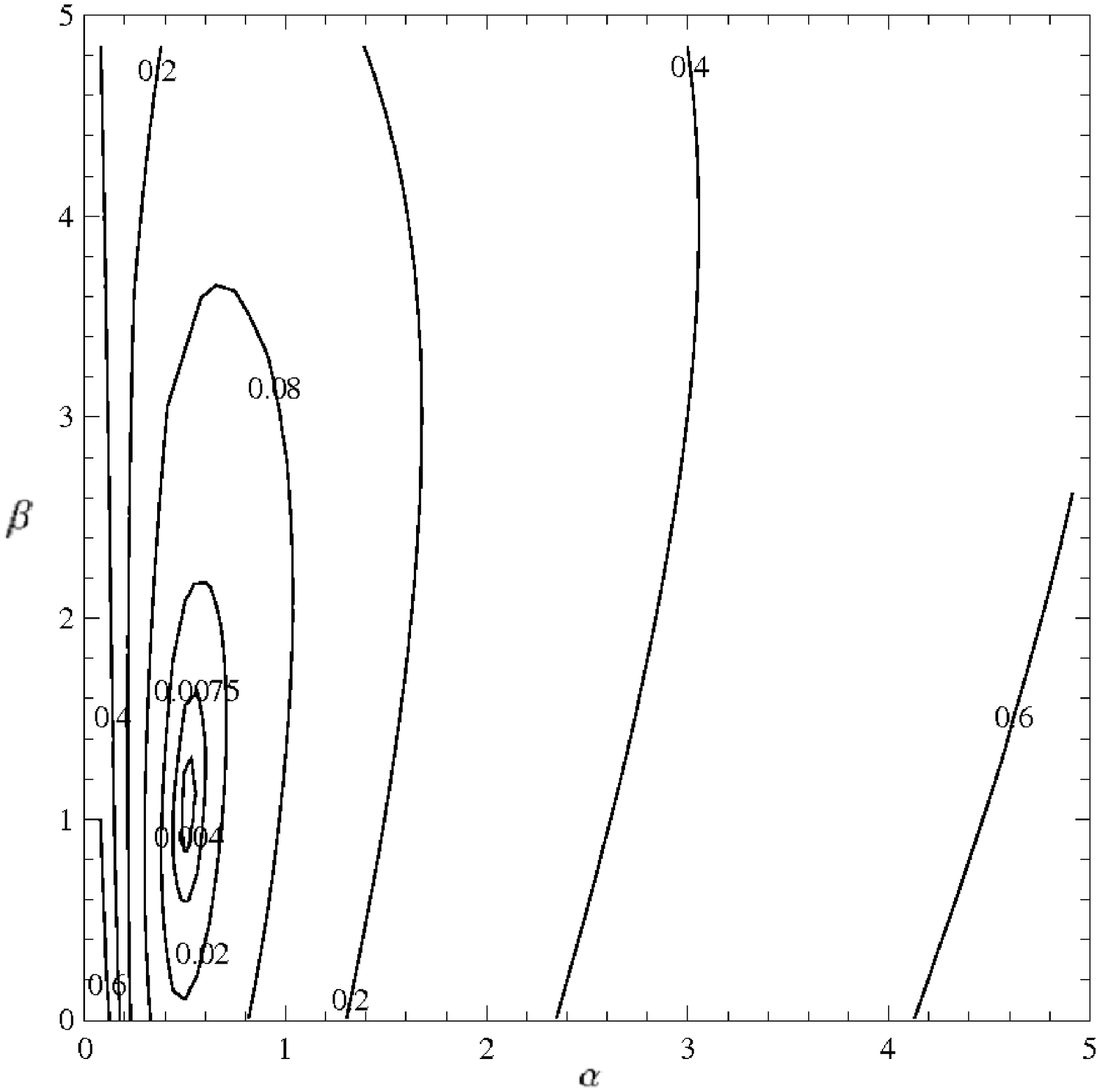}
      \caption{Contours of the function $\Xi$ for the self-consistent decomposition of the rotation curve defined by Eq.~(\ref{paramprof}) with  $\tau_{s}=2.3$.}
         \label{nnpl}
   \end{figure}

\begin{figure}
   \centering
\includegraphics[width=8.7cm, angle=0]{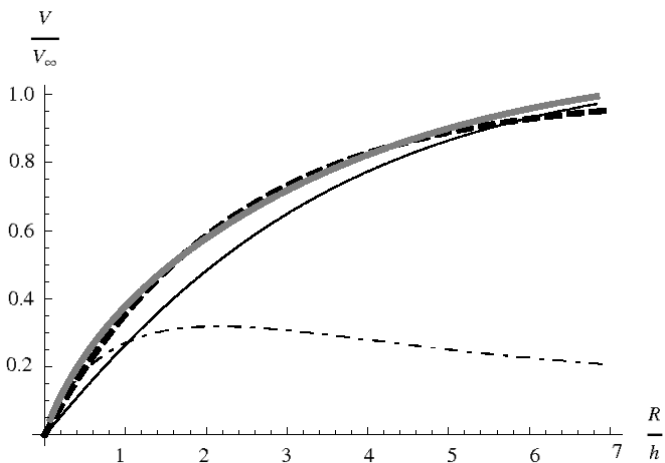}
      \caption{Disk-halo decomposition associated with
      the best self-consistent fit of the case $\tau_{s}=2.3$. The coding is the same as in Fig.~\ref{bfrc}.}
         \label{lrc}
   \end{figure}
The results for the slowly rising rotation curve are shown in Fig.~\ref{pl} and Fig.~\ref{nnpl}. The degeneracy pattern is clearly present in the parametric decomposition, but in this case it is less marked with respect to the cases studied earlier in this paper. Here, the parametric fit is ``pushed" in the direction of an insignificant disk, because the observed $R_{\Omega}$ is too small and can only be ascribed to a dominant dark matter halo. On the other hand, the self-consistent decomposition appears to be able to handle also this case (see Fig.~\ref{lrc}).

From the point of view of the quality of the fits, it is apparent that both decomposition methods work best when fitting the case of $\tau\approx 1$, that is for a rotation curve that follows the empirical correlation illustrated in Fig.~\ref{corrfig}. This is the aspect of conspiracy that was introduced and described in Sect.~\ref{empiric}. In this case, the values of the function $\Xi$ corresponding to the best-fitting $(\alpha, \beta)$ pairs are the smallest. In particular, Figure~\ref{hrc} shows that the case of a fast rising rotation curve cannot be properly described without including a central concentrated mass component, such as a bulge; the value of the residuals $\Xi$ in correspondence of the best-fit (see Fig.~\ref{nnph}) is considerably higher if compared for example to the good fit of Figure~\ref{bfrc}. Similarly, for the opposite case of a slowly rising rotation curve, it is the parametric model which is quite unable to account for such a shape of the rotation velocity, with a similar high value for the best-fit residuals (see Fig.~\ref{pl}).


\end{document}